\documentclass[prd,12pt,tightenlines,nofootinbib]{revtex4}

\usepackage[english]{babel}
\usepackage{graphicx}
\usepackage{psfrag}
\usepackage{amsmath}
\usepackage{amssymb}
\usepackage{bbm}
\usepackage{slashed}
\usepackage{verbatim}
\usepackage[hypertex]{hyperref}

\newcommand{\be}{\begin{equation}}
\newcommand{\ee}{\end{equation}}
\newcommand{\bea}{\begin{eqnarray}}
\newcommand{\eea}{\end{eqnarray}}
\newcommand{\bean}{\begin{eqnarray*}}
\newcommand{\eean}{\end{eqnarray*}}

\renewcommand{\b}{\langle}
\newcommand{\ket}{\rangle}
\newcommand{\bra}{\langle}

\newcommand{\irm}{{\rm i}}
\newcommand{\e}{{\rm e}}

\renewcommand{\d}{{\rm d}}
\newcommand{\cl}[1]{{\mathcal #1}}

\newcommand{\vb}[1]{\mathbf{#1}}
\renewcommand{\v}[1]{\vec{#1}}

\newcommand{\sst}{\scriptstyle}

\newcommand{\bC}{\mathbb{C}}
\newcommand{\bR}{\mathbb{R}}

\newcommand{\clG}{\cl{G}}

\newcommand{\clA}{\cl{A}}

\newcommand{\eq}[1]{(\ref{#1})}
\renewcommand{\sec}[1]{sec.\ \ref{#1}}

\newcommand{\fig}[1]{Fig.\ \ref{#1}}

\newcommand{\tr}{{\rm tr}}

\newcommand{\sgn}{\mathrm{sgn}}

\newcommand{\pic}[4]
{
 \begin{figure}
 \begin{center}
 \includegraphics[height=#3]{#4}
 \end{center}
 \caption{\label{#1} #2}
 \end{figure}
}

\newtheorem{theorem}{Theorem}[section]

\newtheorem{proposition}[theorem]{Proposition}

\newcommand{\qed}{\nobreak \ifvmode \relax \else
      \ifdim\lastskip< 1 em \hskip-\lastskip
      \hskip1.0em plus0em minus0.5em \fi \nobreak
      \vrule height0.75em width0.75em depth0 em\fi}

\newcommand{\vj}{\v{\jmath}}

\newcommand{\gb}{\vb{g}}

\newcommand{\Xb}{\vb{X}}

\newcommand{\alphat}{\tilde{\alpha}}

\newcommand{\rd}{\mathrm{d}}

\newcommand{\ab}{\overline{a}}
\newcommand{\bb}{\overline{b}}

\newcommand{\Uh}{\hat{U}}

\newcommand{\xib}{\overline{\xib}}

\newcommand{\thetat}{\tilde{\theta}}

\begin{document}

\title{Quantum geometry from phase space reduction}

\author{Florian Conrady}
\email{fconrady@perimeterinstitute.ca}
\affiliation{Perimeter Institute for Theoretical Physics, Waterloo, Ontario, Canada}
\author{Laurent Freidel}
\email{lfreidel@perimeterinstitute.ca}
\affiliation{Perimeter Institute for Theoretical Physics, Waterloo, Ontario, Canada}

\begin{abstract}
In this work we give an explicit isomorphism between the usual  spin network basis and the 
direct quantization of the reduced phase space of tetrahedra.
The main outcome is 
a formula that describes the space of SU(2) invariant states by an integral over coherent states satisfying the closure constraint exactly,
or equivalently, as an integral over the space of classical tetrahedra.
This provides an explicit realization of theorems by Guillemin--Sternberg and Hall that describe 
the commutation of quantization and reduction. 
In the final part of the paper, we use our result to express the FK spin foam model as an integral 
over classical tetrahedra and the asymptotics of the vertex amplitude is determined.
\end{abstract}

\maketitle
\section{Introduction}
Recently  a wealth of important developments has occurred following the proposal of new spin foam models 
for gravity \cite{EPR1, FK,LS2, ELPR}. These models allow the inclusion of a non--trivial Immirzi parameter and
have been shown to satisfy\footnote{at least in the Riemannian sector} two very important consistency requirements which have been eluding us for a long time:
firstly, in the semiclassical limit they are asymptotically equivalent to the 
usual Regge discretisation of gravity, independently of the complexity of the underlying cell complex \cite{CF2}.
Secondly, they have been shown to possess SU(2) spin network states as boundary states \cite{EPR2,CF1}.

We can link these new developments with the evolution of our understanding of the so--called simplicity constraints.
These simplicity constraints state (in any dimension \cite{FKP}) that the 
bivector used to contract the curvature tensor in the Einstein action comes from a frame field.
In a simplicial context this constraint splits into three classes: there are the face simplicity constraints (among bivectors associated to the same face),
 the cross--simplicity constraints (among bivectors associated to faces of the same tetrahedron) and the volume constraints (among opposite bivectors).
All these constraints are quadratic, but the volume constraint, 
because it depends on different tetrahedra, also involves the connection and is therefore extremely hard to quantise.
The first major advance came from the work of Baez, Barrett and Crane \cite{BaezB, BarrettCrane} who showed that it is possible to 
 linearise the volume constraint and replace it by a constraint living only at one tetrahedron. This new constraint is the closure constraint and the discrete analogue of the Gauss law generating gauge transformations.
The second key insight in this direction came from the work of Engle, Pereira and Rovelli \cite{EPR1} who showed that one can linearise the simplicity and cross simplicity constraints, opening the path toward a new way of constructing spin foam models and allowing the incorporation of the Immirzi parameter.

There is, however, one unsatisfactory point in this description: if one looks at a path integral version 
of these models \cite{CF1} one quickly recognizes that the simplicity and cross simplicity constraints are imposed strongly, whereas 
the closure constraint is only imposed weakly as an expression of gauge invariance and becomes true only in the semiclassical limit.
Such a distinction does not occur in the continuum theory where all simplicity constraints are imposed strongly \cite{RocheB}.
This is related to the fact that the theory admits a geometrical interpretation (in terms of metrics) only in the semiclassical regime, which
 raised recent criticism  \cite{TB}.
 The purpose of our work is to solve this problem and to show that the new spin foam model
can, in fact, be written explicitly in terms of a sum of amplitudes, where all simplicity constraints are treated on the same footing, i.e.\ imposed strongly 
in the path integral, and where all the boundary spin networks have a geometrical interpretation even before taking
 the semiclassical limit.

It is also interesting to note that along with the  construction of new models
we are also witnessing a merging of two lines of thoughts 
on spin foam models that developed in parallel for a long time and are now finally intersecting.
One line of thought, which  is more canonical in spirit, can be traced back to the seminal work of Barbieri \cite{Barbieri},
 who realized that spin network states  of loop quantum 
gravity can be understood (and heuristically derived) by applying the quantization 
procedure to a collection of geometric tetrahedra in 3 spatial dimensions.
This important work suggested that spin network states and  quantum gravity may be about quantizing 
geometric structures. This idea was then quickly applied to the problem of
quantization of a geometric simplex in $\mathbb{R}^4$, with a breakthrough result being the construction
of the now famous Barrett--Crane model \cite{BarrettCrane}. 
This is also the line of thought developed by Engle, Livine, Pereira, Rovelli in \cite{ELPR}.
The main problem in this approach is that the geometry associated with the quantum  tetrahedra is fuzzy and cannot be 
resolved sharply due to the noncommutativity of the geometric  ``flux'' operators.
This problem has haunted the field for a long time, since it prevents a priori a sharp coupling between neighboring 
4d building blocks, and that was the main issue with the Barrett-Crane model. 

The second line of thought can be traced back to the work of Reisenberger \cite{Reis1} who proposed to think about 
quantum gravity directly in terms of a path integral approach in which we integrate over classical configurations
living on a 2--dimensional spine and where spin foam models arise from a type of bulk discretization.
This line of reasoning is deeply related to the Plebanski reformulation of gravity as a constraint SO(4) 
BF theory \cite{DePF, ReisenbergerLR}.
In this approach it is quite easy to get a consistent gluing condition, however, it is much less trivial 
to find models which have a natural and simple algebraic  expression in terms of recoupling coefficients and it is even more of a  challenge to 
get SU(2) boundary spin networks as boundary states.

The merging between these two approaches starts with a work of Livine and Speziale \cite{LivineS} who proposed to label 
spin network states not with the usual intertwiner basis, but with ``coherent intertwiners'' that are 
 labelled by four vectors whose norm is fixed to be the area of the faces of the tetrahedron.
This is almost the missing link between the two approaches and it has led to a very efficient and geometrical way
 of deriving the new spin foam models \cite{FK, LS2}.
 
 There is, however, an important caveat. The coherent intertwiner resolves the fuzzyness of the quantum space of a tetrahedron
 by construction, but it does so at a high price: the classical configuration no longer satisfies the closure constraint
 unless one takes the semiclassical limit and it is therefore not geometrical.
 
 The resolution of these problems that we propose here is very simple in spirit: one should use 
 coherent states that are associated to the geometrical configuration (see \cite{RovSpe} for an early attempt)
 and relate these ``geometrical'' states to the usual spin network states or the coherent intertwiner states.
 
In this paper, we give a construction of such states,
 we show that they form an overcomplete basis of the space of four--valent intertwiners and we
 give the explicit isomorphism between this new basis of states, where the closure constraint is imposed strongly, and 
 the coherent intertwiner basis.
 
 The proof of this isomorphism amounts essentially to showing that ``quantisation commutes with reduction'' and utilises heavily the beautiful and seminal
  work of Guillemin and Sternberg \cite{GS}.
  All our proofs and derivations are self--contained and we do not assume any knowledge of the mathematical 
  literature \cite{GSbook, Blau, Roberts, Charles, Hall} on geometrical quantisation which is usually the arena in which this work is presented.
  
  The main formula of the paper is described in section \ref{Main}. It expresses the decomposition of identity 
  in the space of four--valent intertwiners 
  $H_{\vj} =  \left(V^{j_{1}} \otimes \cdots \otimes V^{j_{4}}\right)^{\mathrm{SU(2)}}$ as the following integral 
  over coherent states satisfying the closure constraint, and hence as an integral over the space of classical tetrahedra:
  \be
  \mathbbm{1}_{\vj} = \prod_{i}\rd_{j_{i}}  \int_{P_{\vj}^{(0)}} \prod_{i} d^{2}N_{i} \;
  \delta\left(\sum_{i}j_{i}N_{i}\right)\rho_{\vj}(N_{i})  ||\vj,N_{i}\ket \bra \vj, N_{i}||
  \ee
  $||\vj,N_{i}\ket $ are the invariant intertwiners of Livine \& Speziale \cite{LivineS}. The delta function imposes the closure constraint strongly, so the integral is really an integral over the classical phase space.
   Finally, $\rho_{\vj}(N_{i})$ is a positive SU(2) invariant function
   which will be explicitely determined in eq.\ \eq{meas} and 
   which characterises the isomorphism between the usual spin network basis and this ``classical'' basis.   
  
Once this identity is obtained we apply it to the spin foam models given in \cite{FK}: 
it allows us to formulate the path integral in terms of variables on which the closure constraints are imposed strongly.
These variables are well--suited to analyze the asymptotic large spin behaviour of amplitudes, and we use them here to derive
the asymptotics of the FK vertex amplitude.
Shortly after this work a paper by Barrett et al.\ appeared in which the asymptotics of the vertex amplitude is derived as well \cite{BarrettF}.
  
Our paper is organized as follows: in section \ref{sectet} we describe in detail the phase space of the classical tetrahedron and show that it can be obtained as a symplectic quotient. We also state the classical part of the Guillemin--Sternberg isomorphism, which is a remarkable isomorphism between a constrained phase space 
   divided by the action of the gauge group and the {\it un}constrained phase space divided by the complexification of the group.
Section \ref{coherentstate} provides a detailed discussion of coherent states and their link with geometrical quantisation.
In section \ref{QuantRed} and \ref{Main} we present a ``pedestrian'' proof of the Guillemin--Sternberg theorem for tetrahedra and 
construct explicitly the isomorphism between the new geometrical basis and the coherent intertwiner basis. 
In the last section, we rewrite the FK$\gamma$ spin foam model in terms of the new tetrahedral states and exploit this to derive the asymptotics
of the vertex amplitude.

\section{On the space of shape of tetrahedra}
\label{sectet}

\subsection{The classical phase space of the polygon}
Given a four--tuple $\vec{\jmath}\equiv (j_{1},\cdots j_{4})$ of positive real numbers $j_{i}$ we
consider the Polygon space ${\cal P} _{\vj}$ which is the space of 4--sided polygons in $\mathbb{R}^{3}$
with edge length equal to $j_{i}$ modulo rotations.
This space is non--empty if and only if 
\be
2j_{i} \leq j_{1}+\ldots + j_{4} \quad i=1,\ldots, 4\,.
\ee
It can be written as the symplectic quotient of  $ P_{\vj}\equiv S_{j_{1}}^{2}\times \cdots \times S_{j_{4}}^{2}$ by SU(2),
where $S^{2}_{j}$ is the 2--sphere of radius $j$ and $SU(2)$ acts on it by rotation.
In this paper, we  identify $\mathbb{R}^{3}$ with $ \mathfrak{su(2)}^{*}$,  we represent $ \mathfrak{su(2)}$ as the space of anti-hermitian, traceless
 2 by 2 matrices, and  we identify 
$ \mathfrak{su(2)}$ and  its dual using the metric 
$(X|Y)\equiv -2 \tr(XY)$. In this description 
$S^{2}_{j}$ is a (co)adjoint orbit of  $ \mathfrak{su(2)}$:
\be
S^{2}_{j}=\{X=   j(n \tau_{3} n^{-1}) \in  \mathfrak{su(2)}\;|\; n \in SU(2)\}\,,
\ee
where $\tau_{i} \equiv -\irm \sigma_{i}/2$ and $\sigma_{i}$ denote the Pauli matrices (note that the scalar product is such that $(\tau_{a}|\tau_{b})=\delta_{ab}$).
We use a notation where 
 $X= j N$ denotes a point on $S^{2}_{j}$ and $N\equiv n \tau_{3}n^{-1}$
 denotes the corresponding unit vector.
 
$S^{2}_{j}$ is a symplectic manifold
and the associated Poisson structure\footnote{Given a Poisson bracket we can assign to any 
function $h$ on phase space a Hamiltonian vector field
$   \widehat{X}_{h} \equiv {\{} h, \cdot {\}}$. 
When this Poisson bracket is non degenerate it defines a 
symplectic form $\omega$ by 
$$ i_{\widehat{X}_{h}} \omega = - dh,\quad \mathrm{hence} \quad \omega(\widehat{X}_{h_1},\widehat{X}_{h_{2}})=\{h_{1},h_{2}\}. $$}
is given by 
\be
\{ f , g \} (X) = \epsilon^{abc}X_{a}   \partial_{b}f   \partial_{c}g
\ee
with $X_{b} =(\tau_{b}|X)$.
From this it is obvious that linear functions  $h_{Y}(X)\equiv (Y|X)$ 
 generate the SU(2) action on $S^{2}$. We denote by $\widehat{Y}$ the corresponding vector fields:
 \be\label{haction}
 \{h_{Y},f\} =  \widehat{Y}f,\quad \mathrm{with} \quad 
 \widehat{Y}f(X) = \left.\frac{\rd}{\rd t} f( e^{-t Y} X e^{t Y} )\right|_{t=0}.
 \ee
 The symplectic form on $S^{2}_{j}$ is given 
 by
 \be
 \omega_{j}(\widehat{Y}_{1}, \widehat{Y}_{2}) = h_{[Y_{1},Y_{2}]} = \{h_{Y_{1}}, h_{Y_{2}}\}.
 \ee
 
The first important feature of the Polygon space  ${\cal P} _{\vj}$ is that it can be written as a 
symplectic quotient.
Namely, let us first define on   $ P_{\vj}=S_{j_{1}}^{2}\times \cdots \times S_{j_{4}}^{2}$ the Lie algebra valued function
\be
H(X_{i})\equiv \sum_{i} X_{i} = \sum_{i} j_{i} N_{i}\,, \quad\mbox{and}\quad H_{Y}(X_{i}) \equiv (Y|H) = \sum_i h_Y(X_i)\,.
\ee
$H_{X}$ is the Hamiltonian\footnote{These functions are referred to as the ``moment map'' in the mathematical literature \cite{GSbook, Mumford}, whereas in physics they are called a system of first class constraints.} generating the diagonal action $\e^X\in\mathrm{SU(2)}$ on $ P_{\vj}$.
We have the isomorphism 
\be
\label{P(0)}
{\cal P} _{\vj} = P^{(0)}_{\vj}/ SU(2),\quad \mathrm{with} \quad P^{(0)}_{\vj}\equiv H^{-1}(0)\,.
\ee
 This quotient is non singular iff $j_{1} \pm j_{2} \pm j_{3}\pm j_{4}\neq 0$ for any choice of sign.
 When this inequality is satisfied, the action of SO(3) on $H^{-1}(0)$ 
 possesses no fixed point and the quotient is a Haussdorf manifold \cite{Kapo}.

It follows from this isomorphism that ${\cal P} _{\vj}$ is a 2--dimensional symplectic manifold\footnote{ There is a unique symplectic structure ${\Omega}$ on $P^{(0)}_{\vj}/ SU(2)$ such that $ i^{*}(\sum_{i}\omega_{j_{i}}) = \pi^{*}{\Omega}$  where $i :P^{(0)}_{\vj} \to P_{\vj}$ is the inclusion map 
 and $\pi: P^{(0)}_{\vj} \to P^{(0)}_{\vj}/ SU(2)$ is the quotient map.}.
 Symplectic coordinates can be explicitly constructed \cite{Zapata, Kadar, Kapo}:
  one defines $j_{12}\equiv |X_{1}+X_{2}|$, the length of one diagonal of the 4--gon, and 
 $\phi_{12}\in [0,2\pi] $ as the angle between the two oriented triangles $(X_{1},X_{2},X_{1}+X_{2})$ and 
 $(X_{3},X_{4},X_{3}+X_{4})$ which meet along this diagonal.
 Then,
 \be\label{sympcoor}
 \Omega =  \d j_{12}\wedge \d\phi_{12}\,.
 \ee
 To show this, one computes the Hamiltonian flow generated by $j_{12}$, which
 is the ``bending'' flow generating rotations around the diagonal, i.e.\
 \be
 \{j_{12}, f\}(X)=\left.\frac{\rd}{\rd t} f( e^{-t N_{12}} X e^{t N_{12}} )\right|_{t=0}\,,
 \quad 
 \quad
 \mathrm{with}\quad \quad 
 N_{12}\equiv \frac{X_{1}+X_{2}}{|X_{1}+X_{2}|}\,.
 \ee
 This result implies that $\{j_{12},\phi_{12}\}=1$, hence the form of the symplectic structure.

The next step is to demonstrate that the polygon space is a complex K\"{a}hler manifold
and we first review some aspects of complex geometry on the sphere.
The complex structure on the sphere is a map $J$ such that 
$J(\partial_{z})=i\partial_{z}$,  $J(\partial_{\bar{z}})=-i\partial_{\bar{z}}$, where 
the complex coordinates $z$ are defined by
stereographic projection\footnote{Explicitly this gives
\be 
N(z) = \left( \frac{1-|z|^{2}}{1+|z|^{2}} \tau_{3} - \frac{z}{1+|z|^{2}} \tau_{+} - \frac{\bar{z}}{1+|z|^{2}} \tau_{-}\right),\quad \quad
\tau_{\pm} = (\tau_{1}\pm i \tau_{2}).
\ee
The $\tau$ algebra is $$[\tau_{1},\tau_{2}]=\tau_{3}\quad \mathrm{or}\quad  
[\tau_{+},\tau_{-}] =  \frac{2\tau_{3}}{i},\quad\quad [\tau_{3},\tau_{\pm}] =\frac{\pm \tau_{\pm}}{ i} .$$}:
\be \label{ndef}
N(z)= n(z) \tau_{3} n(z)^{-1},
\quad \quad n(z) \equiv\frac{1}{\sqrt{1+|z|^{2}}}\left( \begin{array}{lr}1 &z \\
-\bar{z} &1 \end{array}\right).
\ee
The symplectic structure on the  sphere $S^{2}_{j}$ can be 
expressed in terms of these coordinates as
\be\label{sympmeas}
 \omega_{j}=  \frac{2j}{i}\frac{ \rd z\wedge \rd \bar{z}}{(1+|z|^{2})^{2}}\,,
\ee
its normalisation being such that $\int { \omega}_{j} =-{4\pi j}$.
We can also determine the explicit expressions of the Hamiltonians 
$h_{a}(z)\equiv h_{\tau_{a}}(X(z))$ and the corresponding vector fields\footnote{Observe that $i_{X} (\rd z\wedge \rd \bar{z}) = X^{z}\rd \bar{z}- X^{\bar{z}} \rd z $.} $\widehat{X}_{a}$  in these coordinates:
\bea
h_{{3}}(z) = j \frac{1-|z|^{2}}{1+|z|^{2}}, &\quad \quad &
\widehat{X}_{3} = i\left( z\partial_{z}-\bar{z}\partial_{\bar{z}}\right)\,,\label{ac1}\\
h_{{+}}(z) = j \frac{-2\bar{z}}{1+|z|^{2}}, & &
\widehat{X}_{+} = i\left( \partial_{z}+ \bar{z}^{2}\partial_{\bar{z}}\right)\,,\label{ac2}\\
h_{{-}}(z) = j \frac{-2{z}}{1+|z|^{2}}, & &
\widehat{X}_{-} = -i\left(  {z}^{2}\partial_{{z}}+\partial_{\bar{z}} \right)\,.\label{ac3}
\eea

The existence of a complex structure $J$ implies that we can canonically extend the SU(2) action 
to an action of  SL(2,$\mathbb{C}$) on $S^{2}$. 
That is, we have a mapping between the complexification of $\mathfrak{su(2)}$ and the space of 
{ \it real} vector fields on $S^{2}$ given by $Y_{1} + \irm Y_{2} \mapsto \widehat{Y}_{1} + J(\widehat{Y}_{2})$.
It can be verified that this map is a Lie algebra morphism.
The complex structure determines also a positive definite  metric 
\be
g_{j}(\widehat{Y}_{1},\widehat{Y}_{2}) \equiv \omega_{j}(J(\widehat{Y}_{1}),\widehat{Y}_{2})\,.
\ee
Explicitely this metric is 
\be
g_{j}(\widehat{Y}_{1},\widehat{Y}_{2}) = j\left\{(Y_{1}|Y_{2}) - (Y_{1}|N)(N|Y_{2}) \right\}\,.
\ee

We can  integrate the infinitesimal SU(2) action (\ref{ac1},\ref{ac2},\ref{ac3}) to a group action
$e^{\widehat{Y}}\cdot f(N) = f(e^{-Y} N e^{Y})$ on functions.
Defining $  N(z^{g})\equiv gN(z)g^{-1}$ 
one obtains that 
\be
z^{g} \equiv \frac{az+b}{\bar{a}-\bar{b}z}\quad 
\quad 
{\mathrm{with}} \quad 
\quad 
g= \left( \begin{array}{lr}a &b \\
-\bar{b} & \bar{a} \end{array}\right).
\ee
 This is most naturally seen using the correspondence between the adjoint action on $N(z)$ 
and  the left multiplication on $n(z)$:
\be\label{gn}
g n(z) = n(z^{g}) \left( \begin{array}{lr}\frac{a-b\bar{z}}{|a-b\bar{z}|} &0 \\
0 &\frac{\bar{a}-\bar{b}{z}}{|a-b\bar{z}|} \end{array}\right).
\ee
Thus, the SL(2,$\mathbb{C}$) 
action  on $S^{2}$
is given by fractional transformations
\be\label{frac}
N(z)\mapsto N(z^{g}),\quad\mbox{where}  \quad z^{g}= \frac{az+b}{cz + d}\quad\mbox{and}\quad
g= \left( \begin{array}{lr}a &b \\
c & d\end{array}\right)\,.
\ee
The SL(2,$\mathbb{C}$) action on functions is $g \cdot f(z) = f(z^{g^{-1}}).$

We can now come back to the space of 4--gons and state the first main result that follows from the deep and beautiful work of  
Guillemin and Sternberg  \cite{GS}.
We have just shown that one has a natural action of  SL(2,$\mathbb{C}$) 
on $P_{\vj}$.
As we have seen, the  SU(2) subgroup preserves the constrained  set $P^{(0)}_{\vj}$. On the contrary, the flow generated by the ``imaginary''
elements $J(\widehat{Y})$ is transverse to it. This follows from 
\be
J(\widehat{Y}) h_{Y}(X) = -\omega_{j}(\widehat{Y},J(\widehat{Y}) )= g_{j}(\widehat{Y},\widehat{Y}) >0\,.
\ee
This suggests that we can relax the constraint as long as we ask for invariance under the complexified group
SL(2,$\mathbb{C}$) and this is the essence of the classical part of Guillemin and Sternberg's theorem.
Namely, if one defines 
\be
P^{s}_{\vj}\equiv \{ (X(z_{1}),\ldots, X(z_{4}))\in  P^{(0)}_{\vj} \;|\; z_{i}\neq z_{j}\;\forall\, i\neq j \}\,,
\ee
this is a subset of the unconstrained phase space called the ``stable'' set.
The complement of this set is the set of degenerate polygons and it is of measure $0$.

Then, we have the non--trivial isomorphism \cite{GS,Kapo}
\be \label{GSiso}
I:{\cal P} _{\vj}=P^{(0)}_{\vj}/\mathrm{SU(2)} \to P^{s}_{\vj}/ \mathrm{SL(2,\mathbb{C})}\,.
\ee
This isomorphism is quite surprising at first sight: it implies that we can identify a 
constrained phase space with an {\it un}constrained one provided we 
demand invariance under the complexification of the gauge group.

First, one can show that we have an isomorphism
\bea
P^{s}_{\vj}/ \mathrm{SL(2,\mathbb{C})} &\to& \mathbb{C}\backslash \{0,1\}\,, \\
{[}z_{1},\ldots, z_{4}{]} &\mapsto& Z \equiv \frac{z_{41}z_{23}}{z_{43}z_{21}}\,, \label{crossr}
\eea
where we have denoted $z_{ij}\equiv z_{i}-z_{j}$, the bracket stands for the equivalence class
under SL(2,$\mathbb{C}$) and $Z$ is the cross-ratio.
This is clear, since  we can construct the fractional transformation
$$
z \mapsto z^{g}=  \frac{(z-z_{1})z_{23}}{(z-z_{3})z_{21}}\,,
$$
which is invertible and maps $(z_{1},\ldots, z_{4})$ to $(0,1,\infty, Z)$, and no fractional transformation except 
 the identity maps $(0,1,\infty, Z)$ to itself. The condition $z_{i} \neq z_{j}$ implies that $ Z\neq 0,1,\infty$. 
 
 The map $I$ is then given by $I(z_{1},\ldots, z_{4})=(0,1,\infty, Z)$.
The proof that this map is surjective, and hence an isomorphism, is non--trivial. 
It follows mainly  from the work of Guillemin and Sternberg, but also from the specific study 
of invariant sections. The proof will be given in section \ref{Main}, since it requires 
more results.

\subsection{The space of shape of tetrahedra}
\label{tet}
  
In this section we recall, following the seminal work of Barbieri \cite{Barbieri}, that  the Polygon space ${\cal P} _{\vj}$ is 
 a suitable completion of the space  of tetrahedra  whose face areas are fixed to be equal to $j_{i}$.
 
 One first needs to start with a subset of $ H^{-1}(0)$ consisting of ``nondegenerate''  configurations, where a degenerate configuration
 of $H^{-1}(0)$  is such that $(X_{1},\ldots, X_{4})$ is of rank at most $2$, or equivalentely, such that $\mathrm{det}(X_{1},X_{2},X_{3})=0$.
 If a configuration is degenerate, this means that all the vectors $X_{i}=j_{i} N(z_{i})$ are orthogonal to a fixed direction.
 Since we are only interested in the polygon space, or the space $P^{(0)}_{\vj}$ modulo $SU(2)$, we can always choose this direction to be represented by $\tau_{3}$. The condition $\mathrm{tr}(\tau_{3}N(z))=0$ is the condition that the $z_{i}$ are real, hence the cross ratio
 $Z$ labelling elements of $H^{-1}(0)/SU(2)$ is real.
 Therefore, any  nondegenerate configuration, which also corresponds to a nondegenerate tetrahedron, as we are about to see, is labelled by a non--real cross--ratio. Thus, if we call the space of nondegenerate tetrahedra modulo rotations $\mathrm{Tet}_{\vj}$, we have
 \be
 \mathrm{Tet}_{\vj} \simeq \mathbb{C} \backslash \mathbb{R}\,.
 \ee
 The fact that this space contains two disconnected components follows from the fact that, given a set of edge lengths, there are two possible tetrahedra that can be constructed and differ by a choice of orientation.
 Real cross ratios can be obtained as a limit of purely complex cross ratios. Such limits correspond to singular tetrahedra
 with zero volume and this 
 shows that ${\cal P} _{\vj}$ is a completion of the space of tetrahedra that includes all types of singular or degenerate ones:
 \be
{\cal P} _{\vj} = \overline{ \mathrm{Tet}}_{\vj}
\ee
 If $\vj$ is singular,  i.e.\  $j_{1}\pm j_{2}\pm j_{3}\pm j_{4}\neq 0 $ for some choice of sign, the closure of nondegenerate configurations in $\mathrm{Tet}_{\vj}$ could 
 also contain  cross ratios with  values $0,1$ or $\infty$.

 In order to show this correspondence let us start from a nondegenerate configuration $(X_{1},\cdots, X_{4}) \in H^{-1}(0)$ 
 and let us define 
 \be 
 \ell^{i} \equiv \frac1{4V(X)} \epsilon^{ijkl} u_{j}{[X_{k},X_{l}]}\,,\quad \quad V(X) =- 2  \mathrm{tr}(X_{1}[X_{2},X_{3}])\,,
 \ee
 where $V(X)$ is proportional to the square volume of the tetrahedron\footnote{If one uses the $\mathbb{R}^{3}$ notation for $X$ instead of 
 $\mathfrak{su(2)}$, we have $V= \mathrm{det}(X_{1},X_{2},X_{3})$.}, $u_{i}\equiv \frac12 (1,1,1,1)$ is an auxiliary unit 4-vector whose components are all equal
 and $\epsilon$ is the totally antisymmetric tensor with four indices.
 These vectors satisfy a closure condition and a duality relation with $X$:
 \be
  \sum_{i=1}^{4}\ell^{i}=0\,,\quad \quad  (\ell^{i} |X_{j}) = \delta_{j}^{i} - u^{i}u_{j}\,.
  \ee  
 We can recover $X_{i}$ from  these vectors  by a dual relation
  \be\label{dual2}
  X_{i} = \frac{1}{4V(\ell)} \epsilon_{ijkl} u^{j}{[\ell^{k},\ell^{l}]}\,, \quad\quad V(X) V(\ell) = \frac{1}{4}\,,
  \ee
  with $V(\ell) = \mathrm{det}(\ell_{1},\ell_{2},\ell_3)$.
We can  therefore construct a tetrahedron whose 
  area normal vectors are exactly given by the $X_{i}$. The edge vectors of this tetrahedron are 
  \be\label{edgelength}
  L^{ij} \equiv \sqrt{|V(X)|}(\ell^{i}-\ell^{j})\,.
  \ee
  That is, $X_{i}$ is the vector normal to the face opposite to the vertex $i$ and its norm is twice the area of this face\footnote{A priori, there are two tetrahedra of opposite orientation that have the same area normal vectors $X_i$. With definition \eq{edgelength} we choose one of these two tetrahedra: namely, 
the one for which $\sgn\,\mathrm{det}(L^{14},L^{24},L^{34}) = \sgn\, V(X)$.}.
The relationship between the edge vectors $L^{ij}$ and the $X_{i}$ is
\be
X_{1}= \epsilon [L^{24},L^{34}]\;\; \mbox{etc.\,,}\quad\mbox{where
$V(L) \equiv \mathrm{det}(L^{14},L^{24},L^{34}) = \epsilon \sqrt{|V(X)|}$,}
\ee
and $\epsilon = \sgn\, V(X)$ characterises the orientation of the tetrahedron.
  The proof of these statements can be obtained by direct computation and a 4--dimensional analogue is given in the appendix of 
  \cite{CF2}.
  The relations between the volumes shows that when $X_{i}$ is a sequence of nondegenerate configurations
  converging toward a degenerate one, then  $V(X) \to 0$ thus the volume of the tetrahedra $V(L)$ also goes to $0$.
  Note that at the same time  $V(\ell)\to \infty$.
 
 To complete this geometrical study it is interesting to translate the canonical variables
and the cross ratio into the tetrahedral language.
In order to do so we denote by $\theta_{ij}$ the dihedral angle 
of the edge shared by the face $i$ and $j$, i.e.\ $\cos \theta_{ij} = (N_{i}|N_{j})$.
The first canonical variable $j_{12}$ is related to the dihedral angle $\theta_{12}$ by
\be
j_{12}^{2} = j_{1}^{2}+ j_{2}^{2} + 2 \cos \theta_{12}\,,
\ee
while the ``bending'' angle $\phi_{12}$ is the angle between opposite edges $(12)$ and $(34)$, i.e.\
\be
\cos \phi_{12} = \frac{(\ell^{12}|\ell^{34})}{|\ell^{12}||\ell^{34}|}\,.
\ee
More generally, let us denote by $\phi_{jk}^{i}$ the angle between the edges $(ij)$ and $(ik)$, that is, 
$\cos \phi_{jk}^{i} = (\ell^{ij}|\ell^{ik})/ |\ell^{ij}||\ell^{ik}|$.
We can express the cross ratio, which parametrises all the possible tetrahedral shapes, 
in terms of these geometrical data:
\be
Z= \frac{\sin\frac{\theta_{41}}{2}\sin\frac{\theta_{23}}{2}}{\sin\frac{\theta_{43}}{2}\sin\frac{\theta_{21}}{2}}\, \e^{i(\phi^{4}_{13} + \phi^{2}_{13}- \phi^{1}_{24}- \phi^{3}_{24})}\,.
\ee

\section{The quantum geometry of coherent states}
\label{coherentstate}

In this section we recall some definitions of coherent states and their relations with geometric quantisation.
We refer the reader to the textbook \cite{Perelomov} for more information on these states.

The Lie algebra generators of SU(2) are denoted $J_{i}$ and satisfy the algebra
\be
[J_{+},J_{-}] = 2J_{3}\,,\quad \quad [J_{3},J_{\pm}] = \pm  J_{\pm}\,,
\ee 
and the reality conditions $J_{+}^{\dagger} =J_{-}$, $J_{3}^{\dagger}=J_{3}$.
 The spin $j$ representation of this algebra is 
\bea
T^{j}(J_{+}) |j,m\ket &=&\sqrt{(j+m+1)(j-m)} |j,m+1\ket\,, \nonumber\\
T^{j}(J_{3}) |j,m\ket &=&m  |j,m\ket\,, \nonumber \\
T^{j}(J_{-}) |j,m\ket &=&\sqrt{(j-m+1)(j+m)} |j,m-1\ket\,, 
\eea
with Casimir
$J^{2}=j(j+1)$ and we denote by $V^{j}$ the corresponding carrier space.
Note that in the spin $1/2$ representation we have $J_{a}= \sigma_{a}/2= i\tau_{a}$.
Given such a representation, we define the group element $n(z)$ as in eq.\ \eq{ndef}.
In terms of Lie algebra elements $n(z)$ can be expressed as
\be
n(z)= \e^{zJ_{+}-\bar{z} J_{-} }\,. 
\ee
The group coherent states are defined by
\be
|j,n(z)\ket \equiv T^{j}(n(z)) |j,-j\ket\,.
\ee
In the following section it will be convenient to label coherent states not only in terms 
of group elements $n$, but also directly by the vector $N$ in $S^{2}$ which $n$ represents. That is, we define
a coherent state labelled by $N$:
\be
| j, N\ket \equiv | j, n(N)\ket\,,\quad {\mathrm{with}} \quad n(N) = \frac{1}{\sqrt{2(1 + N_3)}}
 \left( \begin{array}{lr}{1 + N_3} &-{N_1 - i N_2}\\
{N_1 + i N_2} &{1 + N_3}  \end{array}\right)
\ee
In this way, $| j, N(z)\ket =|j,n(z)\ket$.

One key property of these states is their coherence under the tensor product,
that is,
\be
|j,n(z)\ket =|n(z)\ket^{\otimes 2j},\quad \quad \mbox{where} \quad |n(z)\ket \equiv |1/2, n(z)\ket\,.
\ee
To emphasize this coherence property we will preferably use this notation for the coherent state.
The second key property is the property of holomorphicity:
\be\label{holostate}
|n(z)\ket^{\otimes 2j} = \frac{1}{(1+|z|^{2})^{j}} |z\ket^{\otimes 2j }\,,\quad\quad \mbox{where} \quad |z\ket \equiv  e^{z J_{+}} |1/2,-1/2\ket\,.
\ee
From this one can compute the hermitian product of such states,
\be
{}^{\otimes 2j}\bra n(w) |n(z)\ket^{\otimes 2j} = 
\bra n(w) |n(z)\ket^{2j}  = \frac{(1 + \bar{w}z)^{2j}}{(1+|w|^{2})^{j}(1+|z|^{2})^{j}}\,,
\ee
and check the completeness relation 
\be 
\mathbbm{1}_{j} = \frac{\rd_{j}}{2\pi}\int \frac{ \rd^{2}z }{(1+|z|^{2})^{2}}  \left(|n(z)\ket\bra n(z) |\right)^{\otimes 2j}
={\rd_{j}}\int_{S^{2}} { \rd^{2}n(z) }  \left(|n(z)\ket\bra n(z) |\right)^{\otimes 2j}\,,
\ee
where $\rd_{j}= 2j+1$ and $ \rd^{2}n(z)$ is the normalised measure on the unit sphere.
From the definition of the coherent state and the behavior of $n(z)$ under left multiplication expressed in (\ref{gn})\footnote{Note that we can write this relation as $gn(z) = n(z^{g}) \left( \frac{\bar{a}-\bar{b}z}{|\bar{a}-\bar{b}z|}\right)^{-2J_{3}}$.} we can get 
the expression of the SU(2) group action on the group coherent state: 
$$ \nonumber
T^{j}(g)|n(z)\ket^{\otimes 2j} =  |gn(z)\ket^{\otimes 2j} = \left( \frac{\bar{a}-\bar{b}z}{|\bar{a}-\bar{b}z|}\right)^{2j}  |n(z^{g})\ket^{\otimes 2j},\quad z^{g}\equiv \frac{az+b}{-\bb z+ \ab}\quad\mathrm{if}\quad g= \left( \begin{array}{lr}a &b \\
-\bar{b} & \bar{a} \end{array}\right)\,.
$$
Using that ${|\bar{a}-\bar{b}z|^{2}}(1+|z^{g}|^{2}) = (1+|z|^{2})$ if $g \in$ SU(2), we can get the
 corresponding action for the  holomorphic state $|z\ket$ which reads 
\be
T^{j}(g)|z\ket^{\otimes 2j} =   \left({\bar{a}-\bar{b}z}\right)^{2j}  |z^{g}\ket^{\otimes 2j}\,.
\ee
The essential point is that this action is purely holomorphic, since $\partial_{\bar{z}}  |z\ket^{\otimes 2j}=0$.
Therefore, as in the classical setting, we can extend the SU(2) action to an action of SL(2,$\mathbb{C}$).
That is,
\be \label{holog}
T^{j}(g)|z\ket^{\otimes 2j} =   \left({cz +d}\right)^{2j}  |z^{g}\ket^{\otimes 2j},\quad z^{g}\equiv \frac{az+b}{cz+d}\quad\mathrm{if}\quad
g= \left( \begin{array}{lr}a &b \\
c & d \end{array}\right)\in \mathrm{SL(2,}\mathbb{C})\,.
\ee
We can lift this action back from the holomorphic to the group coherent state and finally get 
\bea
T^{j}(g)|n(z)\ket^{\otimes 2j} &=&   \left(\frac{cz +d}{|cz +d|}\right)^{2j} \left(\frac{|cz +d|^{2} (1+ |z^{g}|^{2})}{1+|z|^{2}}\right)^{j}  |n(z^{g})\ket^{\otimes 2j}\nonumber \\
&=&   \left(\frac{cz +d}{|cz +d|}\right)^{2j} \left(\frac{\bra z|g^{\dagger} g|z\ket}{\bra z|z \ket} \right)^{j}  |n(z^{g})\ket^{\otimes 2j}\\
&\equiv& \rho_{g}^{-j}(z)  |n(z^{g})\ket^{\otimes 2j}\,. \label{SLnz}
\eea
In the second  line we used $g$ interchangeably for $T^{1/2}(g)$.
We have introduced a pairing $\rho_{g}(z)$ between an element of $  \mathrm{SL(2,}\mathbb{C})$ and an element of $S^{2}$
which is explicitely given by 
\be
\rho_{g}(z)\equiv  \frac{1+|z|^{2}}{|cz +d|^{2} + |az +b|^{2}}    \left(\frac{|cz +d|}{cz +d}\right)^{2},\quad 
\mathrm{for}\quad g= \left( \begin{array}{lr}a &b \\
c & d \end{array}\right)\in \mathrm{SL(2,}\mathbb{C})\,.
\ee
 
From (\ref{holog}) one can derive  explicitly the action of infinitesimal 
SU(2) generators acting on  the holomorphic states:
\bea \label{rep}
\bra s |T^{j}(J_{+}) |z\ket^{\otimes 2j} &=& \partial_{z}  \bra s |z\ket^{\otimes 2j}\,, \nonumber\\
\bra s |T^{j}(J_{3})  |z\ket^{\otimes 2j} &=& (z\partial_{z} - j  )\bra s |z\ket^{\otimes 2j}\,, \nonumber \\
\bra s |T^{j}(J_{-}) |z\ket^{\otimes 2j} &=&  (-z^{2}\partial_{z} +2 j z ) \bra s |z\ket^{\otimes 2j}\,.
\eea
Here, $|s\ket$ is any state in the spin $j$ representation.
From this we can now easily compute the expectation value of Lie algebra 
generators and find that it is simply given by the moment map $h_{a}$:
\be
\label{hv}
\frac{\bra j,z |T^{j}(J_{a})  |j,z\ket}{\bra j,z |j,z\ket} =  -j N_{a}(z) = - h_{a}(z)
\ee
We will also need the expression for the action of a general vector field $\widehat{X}$
on the moment map: for $X \in  \mathfrak{sl(2 ,\mathbb{C})}$ and $Y \in  \mathfrak{su(2)}$
\be\label{Gaction}
\e^{-\widehat{X} }\cdot h_{Y}(z) = \frac1{i}
\frac{\bra j,z |T^{j}( e^{X^{\dagger}} Y e^{X}) |j,z\ket}{\bra j,z |T^{j}( e^{X^{\dagger}} e^{X})  |j,z\ket}\,.
\ee
This follows directly from the fact that on the one hand $e^{-\widehat{X} }\cdot h_{Y}(z)= h_{Y}(z^{g})$ with $g=e^{X}$,
and on the other hand $ h_{Y}(z^{g})$ can be expressed as the right--hand side of (\ref{Gaction}) using the coherent state transformation
(\ref{holog}) and the relation (\ref{hv})
 between the Hamiltonian and the expectation value (recall that $J_{a}=i\tau_{a}$).

We can also see that in  the case $X\in \mathfrak{su(2)}$ (that is, $X$ anti-hermitian) 
the previous expression reduces to the action (\ref{haction}) of SU(2) on functions on $S^{2}$.
By differentiating the previous expression 
we also get a relation  between 
the metric and  the symmetric connected 2--point
function: given $X,Y \in  \mathfrak{su(2)}$  we have 
\bea\label{g}
-g_{j}(\widehat{X},\widehat{Y})  &=& \frac{\bra j,z |T^{j}( X Y+ YX) |j,z\ket}{\bra j,z  |j,z\ket} -2 \frac{\bra j,z |T^{j}(X)|j,z\ket}{\bra j,z  |j,z\ket}\frac{\bra j,z |T^{j}(Y) |j,z\ket}{\bra j,z  |j,z\ket} \nonumber \\
&=& -j X^{a}(\delta_{ab} - N_{a}(z)N_{b}(z)) Y^{b} 
\eea
with $X=X^{a}\tau_{a}$.
More generally, the connected two point function is related to the hermitian form
\bea
-g_{j}(\widehat{X},\widehat{Y}) + \irm\omega_{j}(\widehat{X},\widehat{Y}) &=& 2\frac{\bra j,z |T^{j}( X Y) |j,z\ket}{\bra j,z  |j,z\ket} -2 \frac{\bra j,z |T^{j}(X)|j,z\ket}{\bra j,z  |j,z\ket}\frac{\bra j,z |T^{j}(Y) |j,z\ket}{\bra j,z  |j,z\ket}\\
&=&-j X^{a}\left(\delta_{ab} - N_{a}(z)N_{b}(z) - i \epsilon_{abc} N^{c}(z) \right) Y^{b}\,.  \label{h}
\eea

\subsection{Geometric quantisation}

The previous description of coherent states is the usual one used in the physical literature.
In the mathematical literature \cite{GSbook, Roberts, Charles} one uses preferably the language of geometric quantisation to describe the same 
construction.
For the reader's convenience and in order to connect the mathematical and physical terminology
we describe the relationship between these two languages in a pedestrian manner.

Firstly, in order to define the geometrical quantisation we need a Hermitian line bundle $L$ over 
$S^{2}$ of curvature $i \omega_{j}$.
A natural hermitian line bundle over $S^{2}$ is given by $\mathbb{CP}_{1}\to S^{2}$. Denoting this  line bundle by $L$, 
all the other line bundles are obtained  by tensorisation $L_{j} \equiv L^{\otimes 2j}$.
The conditions on this line bundle mean that we have a covariant derivative $\nabla = d +A $, 
with hermitian connection and curvature given by
\bea
A_{z} =-j\frac{\bar{z}}{1+|z|^{2}}\,, \quad 
A_{\bar{z}} = j\frac{z}{1+|z|^{2}}\,, \\
F_{z\bar{z}} = \partial_{z}A_{\bar{z}}-\partial_{\bar{z}}A_{z}= \frac{2 j}{(1+|z|^{2})^{2}}\,.
\eea
In geometric quantisation one identifies the representation space of 
spin $j$ with the space of holomorphic sections of $L_j$ denoted $H^{0}(S^{2},L_j)$. Explicitly,  the space of sections of $L_{j}$  is the space
of functions $s(z,\bar{z})$ such that both $s$ and $\tilde{s}(z) = \frac{z^{2j}}{|z|^{2j}} s(-\frac{1}{z})$ are holomorphic sections. That is, they are  analytic and
 solutions of $\nabla_{\widehat{X} -\irm J(\widehat{X})}s =0$. This implies that 
 $$s(z,\bar{z}) =\frac{\hat{s}(z)}{(1+|z|^{2})^{j}}\,,$$ where $\hat{s}(z)$ is a holomorphic polynomial of degree at most $2j$.
 
 $H^{0}(S^{2},L_j)$ carries a representation of SU(2) given by the differential operators 
 \be
 Y\cdot s = (\nabla_{\widehat{Y}} -i h_{Y}) s
 \ee
 where $Y$ is an anti--hermitian element and $h_{Y}$  is the momentum map associated with the SU(2) action on $S^{2}_{j}$.
 The fact that this forms a representation follows directly from 
 the symplectic identities 
 \be 
 \omega(\widehat{X},\widehat{Y}) = -\irm F(\widehat{X},\widehat{Y}) = \widehat{X} h_{Y} = h_{[X,Y]}\,.
 \ee
 The link between this representation and the previous coherent state representation can be described as follows.
 Suppose that $|s\ket$ is a state in $V_{j}$ and define 
 \be
 s(z)\equiv \bra s |n(z)\ket^{\otimes 2j}\,.
 \ee
 One can verify that this is a holomorphic section of $L_j$. In particular,  $\nabla_{\bar z}s(z) =0$,
 which follows from 
 \be
 \nabla_{\bar z}s(z) = \frac{\partial_{\bar{z}}\hat{s}(z)}{(1+|z|^{2})^{j}}=0\,,\qquad 
 \hat{s}(z) \equiv \bra s | j, z\ket\,.
 \ee
 Furthermore, the geometric representations coincide with the group coherent state representation:
 \be
 Y\cdot s (z) = \left(\nabla_{\widehat{Y}} - \irm h_{Y}\right) s(z) = -\bra s |T^{j}(Y)| n(z)\ket^{\otimes 2j}=-\frac{\bra s |T^{j}(Y)|z \ket^{\otimes 2j}}{\bra z| z\ket^{\otimes 2j}}
 \ee
 This follows from the fact that the ``Lagrangians ''  $L_{X}\equiv \theta(\widehat{X}) - h_{X}$ associated with the 
holomorphic  symplectic potential $\theta \equiv 2\irm j  \frac{\bar{z} \rd z }{1+|z|^{2}}$ coincide with the scalar terms in \eq{rep},
that is, 
\be
L_{X_{3}} = -j,\quad L_{X_{+}}=0,\quad L_{X_{-}}=2jz\,.
\ee

 The language of geometric quantisation explains also the extension of the SU(2) action to SL(2,$\mathbb{C}$).
 Firstly, the complex structure allows us to extend the mapping $X \to \widehat{X}$ from Lie algebra to real vector fields 
 to hermitian operators by 
 $\widehat{(\irm X)} = J(\widehat{X})$. Then,  the holomorphicity of the section implies that $\nabla_{J(\widehat{X})} =  \irm \nabla_{\widehat{X}}$,
 hence
 \be
 (\irm Y) \cdot s (z) = \left(\nabla_{\widehat{(\irm Y)}} - i h_{\irm Y}\right) s(z)= \left(\nabla_{J(\widehat{Y})} + h_{Y}\right) s(z)\,.
\ee

\section{Quantisation commutes with reduction }
\label{QuantRed}

Now that we have described all the necessary ingredients we can 
focus on our main task: the construction of the quantisation of the space of shape of tetrahedra, which is the 4--gon space ${\cal P} _{\vj}$.
One route to the quantisation of this space is the route proposed by Barbieri \cite{Barbieri} in his seminal work and studied since then in 
the quantum gravity literature: 
according to this scheme one first quantises the phase space of four vectors $X_{i}\in S^{2}_{j_{i}}$, promotes these vectors to operators associated to 4 copies of $\mathfrak{su(2)}$, where in each case $X_{i}^{2}= j_{i}(j_{i}+1)$, and {\it then } one imposes the constraints $\sum_{i} X_{i} =0$ at the operatorial level.
The Hilbert space for this quantisation 
  is the space of 4-valent intertwiners 
\be 
H_{\vj} =\left(V^{j_{1}}\cdots \otimes V^{j_{4}}\right)^{\mathrm{SU(2)}}\,,
\ee
i.e.\ the space of $SU(2)$ invariant vectors.
  A basis for this space is given by invariant states $|j_{i}, i_{12}\ket$ diagonalising the operators $(X_{1}+ X_{2})^{2}$ with eigenvalue $i_{12}(i_{12}+1)$
  and labelled by 
  an intertwiner label $i_{12}$ which is a spin. The completeness relation for these states is $ \mathbbm{1}_{\vec{\jmath}}= \sum_{i}\rd_{i} |j_{i}, i\ket\bra j_{i}, i|$.
  Recently, it has been advocated by Livine and Speziale \cite{LivineS} that a more convenient basis is the 
  ``coherent intertwiner'' basis. For this basis one starts with a product of 4 coherent states and then averages them with respect to the group action in order to get an invariant state:
  \be\label{ginvcs}
  ||\vec{\jmath},N_{i}\ket\equiv  \int_{\mathrm{SU(2)}} \!\! \rd k  \, \otimes_{j_{i}} T^{j_{i}}(k)  |j_{i},N_{i}\ket
  \ee
  In this basis the resolution of identity reads 
  $$ \mathbbm{1}_{\vec{\jmath}}= \prod_{i}\rd_{j_{i}}\int\prod_{i}\rd^{2}N_{i}\; ||\vec{\jmath}, N_{i}\ket \bra \vec{\jmath},N_{i}||\,,$$
  where $\mathbbm{1}_{\vec{\jmath}}$ denotes the identity in (or projector onto) $H_{\vj}$ and $\d^2 N$ is the normalised measure on $S^{2}$.
  
  Another route toward the quantisation of the same space, which is conceptually much simpler, is to 
  first impose the constraints and then quantise. As we have seen, the reduced space  ${\cal P} _{\vj}= P^{(0)}_{\vj}/\mathrm{SU(2)}$ is 
  a symplectic manifold and we even know some nice symplectic coordinates $(j_{12},\phi_{12})$ 
  on this space (see (\ref{sympcoor})).
   All SU(2) invariant function can be expressed as a function of these variables and the quantisation in this real polarisation 
 is straigthforward:
 the Hilbert space ${H}_{\vec{\jmath}}^{\mathrm{(0)}}$ is the space of $L^{2}$ function of $\phi_{12}$ (say) and the completeness relation reads  $ \mathbbm{1}_{\vec{\jmath}}=  \int_{0}^{2\pi} \rd \phi_{12}  |\phi_{12}\ket\bra \phi_{12}| $.
 
 We have also seen that ${\cal P}_{\vj}$, being the quotient of a K{\"a}hler manifold\footnote{ Recall that a K{\"a}hler manifold is a symplectic manifold with a complex structure compatible with the symplectic form, i.e.\ $\omega(\widehat{X},\widehat{Y}) =\omega(J(\widehat{X}),J(\widehat{Y}))$, and $B(\widehat{X},\widehat{Y})=\omega(J(\widehat{X}),\widehat{Y})$ is a positive definite pairing.},
is itself a K{\"a}hler manifold. The complex coordinate associated with this  K{\"a}hler structure is the cross ratio $Z(z_{i})$ (see eq.\ \eq{crossr}).
Following the procedure of geometric quantisation presented in the previous section, we can construct the coherent states $|Z\ket$ associated to this 
complex coordinate. The completeness relation for these states is
$$
\mathbbm{1}_{\vec{\jmath}} =  \int_{{\cal P}_{\vec{\jmath}}} {\Omega}(Z)\; |Z\ket\bra Z|
$$
with $\Omega$ being the symplectic structure on ${\cal P}_{\vec{\jmath}}$.

The key and essential result following from the general framework of Guillemin--Sternberg \cite{GS} and the detailed study of its unitarity properties by B.\ Hall \cite{Hall} is that one should expect  an isomorphism
  between  the two Hilbert spaces $\hat{I}:{H}_{\vec{\jmath}}^{\mathrm{(0)}} \to {H}_{\vec{\jmath}}$.
 This isomorphism is the mathematical translation of  the statement that
 ``Quantisation commutes with reduction''.
 The purpose of section \ref{Main} is to give a proof of this isomorphism {\it and } to construct it explicitly.
 As we will see, the correspondence is such that  the ``reduced'' coherent state $|Z\ket$ maps, up to an overall normalisation, 
 to a group--averaged coherent state, labelled by a point on the constraint surface:
\be
I(|Z\ket) = \tilde{\rho}_{\vj}(z_{i})\, ||\vec{\jmath}, N_{i}\ket\quad\mbox{for all $z_i$ s.t.\ $Z = Z(z_i)$ and $\sum_i j_i N(z_i) = 0$\,.}
\ee
$\tilde{\rho}_{\vj}(z_{i})$ is a strictly positive function invariant under the action of SL(2,$\mathbb{C}$). 
The proof we are going to give is self--contained and does not require any prior knowledge of geometric quantisation and the work of Guillemin--Sternberg.

\subsection{3-- and 4--valent section}
Before going into the proof of our main statement we first describe in more detail the space of 
3-- and 4--valent invariant sections and some properties of the SU(2) invariant group coherent state (\ref{ginvcs}).
Starting  from this state it will be convenient to 
define the corresponding holomorphic state
\be
 {||\vj,z_{i}\ket} \equiv{\prod_{i}(1+|z_{i}|^{2})^{j_{i}}}||\vj, N(z_{i})\ket\,.
\ee
Given a state $|S\ket$ in $ H_{j} =\left(V^{j_{1}}\otimes \cdots \otimes V^{j_{4}}\right)^{\mathrm{SU(2)}}$, we can construct 
the ``invariant holomorphic section''
\be
S^{\vj}(z_{i}) \equiv \bra S || \vj, z_i\ket\,.
\ee
Such sections are characterised by the fact that they are polynomials of degree $2j_{i}$ in the variable $z_{i}$ and 
by their property of invariance under  $\mathrm{SL(2,\mathbb{C})}$ transformations: namely,
 \be\label{SLinv}
  S^{\vj}(z_{i}+a) = S^{\vj}(z_{i})\,,\quad \prod_{i}z^{2j_{i}}_{i} S^{\vj}\left(-\frac1{z_{i}}\right) = S^{\vj}(z_{i})\,,
  \quad S^{\vj}(\lambda Z_{i}) = \lambda^{2j_{i}} S^{\vj}(z_{i})\,.
  \ee
Conversely, any such section defines a state in $ H_{j} =\left(V^{j_{1}}\cdots \otimes V^{j_{4}}\right)^{\mathrm{SU(2)}}$
 by 
\be
|S\ket =\prod_{i}\rd_{j_{i}} \int \prod_{i}\rd^{2}N(z_{i})\, |\vj, N(z_{i})\ket \frac{ \overline{S^{\vj}(z_{i})} }{(1+|z_{i}|^{2})^{j_{i}}}.
\ee
Let us first look at the space of invariant sections which depends only on three entries $(z_{1},z_{2},z_{3})$. The conditions of
$\mathrm{SL(2,\mathbb{C})}$ invariance
(\ref{SLinv}) fix the form of the section uniquely up to a normalisation.
This unique section is  given by
\be
S^{j_{1}j_{2}j_{3}}(z_{1},z_{2},z_{3})= {\sqrt{N_{{j_{1}j_{2}j_{3}}}}}{
(z_{1}-z_{2})^{\Delta_{3}}(z_{2}-z_{3})^{\Delta_{1}}(z_{3}- z_{1})^{\Delta_{2}} }\,,
\ee
where $\Delta_{i} = {j_{1}+j_{2}+j_{3}}-2j_{i}$ and $N_{j_{i}}$ is a normalisation factor chosen such that 
\be
1= \bra S|S\ket = \prod_{i=1}^{3}\rd_{j_{i}} \int \prod_{i=1}^{3}\rd^{2}N(z_{i}) \frac{|S^{j_{1}j_{2}j_{3}}(z_{1},z_{2},z_{3})|^{2}}{{{\prod_{i=1}^3(1+|z_i|^2)^{2j_i}}}}\,. 
\ee
It is explicitely given by (see \cite{VK})
\be
N_{j_{1}j_{2}j_{3}}^{-1}= 
\frac{[j_{1}+j_{2}+j_{3}+1]! [-j_{1}+j_{2}+j_{3}]! [j_{1}-j_{2}+j_{3}]! [j_{1}+j_{2}-j_{3}]! }{[2j_{1}]![2j_{2}]![2j_{3}]!}\,.
\ee
Next we look at the space of invariant sections which depend  on four  entries $(z_{1},z_{2},z_{3},z_{4})$. The condition of
$\mathrm{SL(2,\mathbb{C})}$ invariance no longer fixes the functional form of these sections uniquely.
It implies, however, that they  have the form
  \be\label{4-v}
 S(z_{i}) = z_{12}^{j_{1}+j_{2}-j_{3}+j_{4}} z_{23}^{(-j_{1}+j_{2}+j_{3}-j_{4})}  z_{31}^{(j_{1}-j_{2}+j_{3} -j_{4})} z_{34}^{2j_{4}} 
 s(Z)\,,
 \ee
 where $Z$ is the cross ratio
 \be
 Z\equiv \frac{z_{41}z_{23}}{z_{43}z_{21}}\,,\quad \quad z_{ij}\equiv z_{i}-z_{j}\,,
\ee
and 
$ s(Z) = \lim_{_{X\to \infty}} X^{-2j_{3}} S(0,1,X,Z).$
This section should also be a polynomial in $z_{i}$ of degree at most $2j_{i}$. Therefore, $s$ should be a polynomial in 
$Z$ of degree smaller than min$(2j_{4}, j_{4}+j_{1}+j_{2}-j_{3})$ and of valuation at least 
max$(0,j_{4}-j_{1}+j_{2}-j_{3},j_{4}+j_{1}-j_{2}-j_{3} )$. 

The usual spin network basis  $|\vj, i\ket$ corresponds to 
a section denoted $ S_{i}^{\vj}(z_{i})\equiv \bra \vj,i || \vj, z\ket$.
This section  diagonalises the operator $\Delta_{12}\equiv X^{(1)}\cdot X^{(2)}$
with eigenvalue $ i(i+1) -j_{1}(j_{1}+1) -j_{2}(j_{2}+1)$,
where $X^{(i)}$ denotes the action on the variable $z_{i}$.
This differential  operator is given by 
\be
\Delta_{12}= -\frac12 z_{12}^{2} \partial_{1}\partial_{{2}} + z_{12}(j_{1}\partial_{2}-j_{2}\partial_{1} )
+ j_{1}j_{2}.
\ee
The projection of the coherent state on the spin network basis can also be expressed as an integral
involving the product of two trivalent invariant (Clebsch--Gordan) sections:
\be
S_{i}^{\vj}(z_{i}) = \rd_{i} 
\int   \frac{S^{j_{1}j_{2}i}(z_{1},z_{2},z) S^{ij_{3}j_{4}}(-\bar{z}^{-1},z_{3},z_{4})}{(1+|z|^{2})^{2i}} {z^{2i}} \, \d^{2}N(z)\,.
\ee

\section{The Quantum equivalence }\label{Main}

The purpose of this section is to first present  some of the key properties satisfied by 
 the coherent intertwiner, then to
give a proof of the classical correspondence (\ref{GSiso}) and finally to state and prove our main result: the explicit isomorphism between the 
space of 4--valent intertwiners and the space generated by coherent states labelled by classical tetrahedra.

\subsection{ Some properties of invariant coherent states}
 
As discussed previously, the  coherent intertwiner (\ref{ginvcs}) is
 defined by the projection of the usual group coherent states:
 \be
  ||\vec{\jmath},N_{i}\ket\equiv  \Pi^{(0)} \otimes_{i} |j_{i},N_{i}\ket,\quad \quad
  \Pi^{(0)} \equiv \int_{\mathrm{SU(2)}} \!\! \rd k  \, \otimes_{j_{i}} T^{j_{i}}(k)
  \ee
Here, $\Pi^{(0)}$ is a group averaging projector annihilating the diagonal SU(2) action.
Since the coherent state is holomorphic this invariance extends to the $\mathrm{SL(2},\mathbb{C})$ action.
Together with the rules of transformation (\ref{SLnz}), the invariance property $\Pi^{(0)} \left(\otimes_{i} T^{j_{i}}(g) \right) = \Pi^{(0)}$ implies the following key transformation property: given any $g \in $ SL(2,$\mathbb{C}$)
 \be\label{stateinv}
 ||\vj, N(z_{i}^g)\ket  =\rho^{\vj}_{g}(z_{i}) ||\vj, N(z_{i})\ket\,,
 \ee
 where the prefactor equals the product 
 \be
 \rho^{\vj}_{g}(z_{i}) \equiv \prod_{i} \rho_{g}^{j_{i}}(z_{i})\,,\quad \quad 
 \rho_{g}(z) =  \left(\frac{|cz +d|}{cz +d}\right)^{2} \frac{\bra z|z \ket}{\bra z|g^{\dagger} g|z\ket}\,,
 \ee
 and $\bra z|z \ket = 1+|z|^{2}$.
 We also introduce a notation for the norm of the covariant intertwinner  $||N(z_{i})||_{\vj}^{2}\equiv   \bra \vj, N(z_{i}) ||\vj, N(z_{i})\ket$.
 The factor $\rho_{\vj}$ is just the ratio of this norms
 \be\label{rhoN}
 |\rho_{g}^{\vj}|^{2}(z_{i}) =\frac{||N(z_{i}^{g})||_{\vj}^{2}}{||N(z_{i})||_{\vj}^{2}}.
 \ee
 
 Note that in the following we will also use the notation $\rho_{g}(N_{i})$ with the obvious definition
 $\rho_{g}(N(z_{i}))\equiv \rho_{g}(z_{i})$.
 The prefactor $ \rho^{\vj}_{g}(z_{i})$ controls the behavior of the invariant states along the 
 $\mathrm{SL(2},\mathbb{C})$ orbits and satisfies a number of key properties that we now list:
 \begin{proposition} 
 Suppose that $(z_i)$ is a point in $P_{\vj}$ with nonzero norm $||N(z_{i})||_{\vj}$.
 
 1- The first derivatives of $||N(z_{i})||_{\vj}$ for an anti-hermitian $X= X^{a}\tau_{a}$ 
 are given  by 
 \bea
 \widehat{(iX)}\ln||N(z_{i})||_{\vj} &=& - H_{X}(z_{i})\,,  \label{H} \\
  \widehat{(iY)} \widehat{(iX)}\ln ||N(z_{i})||_{\vj} &=&-G(X,Y)(z_{i})\,,  \label{G}
 \eea
 where $H_{X}(N_{i}) = \sum_{i} j_{i} N_{i}$ is the Hamiltonian constraint and $G=\sum_{i} g_{j_{i}}$ is the metric 
 on the orbits $ \mathrm{SL(2},\mathbb{C}) \cdot (z_{i})$. Explicitly, 
\be G_{ab}\equiv G(\tau_{a},\tau_{b}) =\sum_{i} j_{i} (\delta_{ab}-N_{a}(z_{i})N_{b}(z_{i}) ).\label{Gdef}\ee

2- Suppose that $(N_{i}) \in P^{(0)}_{\vj}$, i.e.\ $ \sum_{i} j_{i} N_{i} =0$, then 
 $|\rho^{\vj}_{g}(N_{i})| \leq 1$  for all $g \in \mathrm{SL(2},\mathbb{C})$. Moreover,  the equality is satisfied 
 iff $g \in \mathrm{SU}(2)$.
 
 3- We can relate the norm of  $\rho_{g}^{\vj}(z_{i})$ to the integral of the Hamiltonian constraint: 
 \be
  |\rho^{\vj}_{e^{iX}}(N_{i})|^{2} = \exp \left(\int_{0}^{1} \mathrm{d} t \; \e^{-t\widehat{iX}}\cdot H_{(X-X^{\dagger})}( z_{i})   \right)\,,\ee
  with $X\in \mathfrak{sl(2,\mathbb{C})}$.
\end{proposition}
Let us start by proving point 3: given $X \in \mathfrak{sl(2,\mathbb{C})}$ we have
\bea
 |\rho^{\vj}_{e^{iX}}(z_{i})|^{2}&=& \prod_{i} \frac{\bra j_{i}, z_{i}|  j_{i}, z_{i}\ket}{\bra j_{i}, z_{i}| e^{-iX^\dagger} e^{iX}|  j_{i}, z_{i}\ket}\\
&=& \exp \left(-\int_{0}^{1} \mathrm{d} t \,\partial_{t} \sum_{i}\ln {\bra j_{i}, z_{i}|  e^{-itX^{\dagger}}  e^{itX}|j_{i}, z_{i}\ket}\right) \\
&=& \exp \left(-\int_{0}^{1} \mathrm{d} t \, \sum_{i} \frac{\bra j_{i}, z_{i}|   e^{-itX^{\dagger}}(-iX^{\dagger}+iX)e^{itX}|j_{i}, z_{i}\ket}{\bra j_{i}, z_{i}| 
e^{-itX^{\dagger}}  e^{itX}|j_{i}, z_{i}\ket}\right) \\
&=& \exp \left(\int_{0}^{1} \mathrm{d} t\; H_{(X-X^{\dagger})}(e^{itX}\cdot z_i)   \right)
=\exp \left(\int_{0}^{1} \mathrm{d} t \; \e^{-tJ(\widehat{X})}\cdot H_{(X-X^{\dagger})}( z_i)   \right)\,.
\eea
In the last equalities we have used the definition of $H_{X}$ in terms of the expectation value given in section \ref{coherentstate}.
This expression coincides with  the general expression derived  by B. Hall \cite{Hall}.
Taking the logarithm of this formula and restricting to $X\in \mathfrak{su(2)} $ one obtains that 
\be\label{rhou}
\ln(|\rho^{\vj}_{e^{iuX}}(z_{i})|) = \int_{0}^{u} \mathrm{d} t\; \e^{-tJ(\widehat{X})}\cdot H_{X}(z_i)\,.
\ee
By taking the derivative of this expression with respect to $u$, one obtains that 
\be\label{firstder}
\partial_{u} \left.\ln(|\rho^{\vj}_{e^{iuX}}(z_{i})|)\right|_{u=0} = H_{X}(z_{i}) = - J(\widehat{X}) \ln||N(z_{i})||_{\vj}\,,
\ee
where the second equality follows from the relationship (\ref{rhoN}):
\be
\ln(|\rho^{\vj}_{e^{\irm u X}}(z_{i})|)  = \left(e^{- u J(\widehat{X})}-1\right) \ln||N(z_{i})||_{\vj}
\ee
By taking  another  derivative of (\ref{firstder})  one obtains easily 
 property 1, since $J(\widehat{X}) H_{Y} = G(X,Y)$.

 From the relation (\ref{stateinv}) one clearly sees that if $||N(z_{i})||_{\vj} \neq 0$, 
 then  for any $g \in \mathrm{SL(2,}\mathbb{C})$ we also have $||N(z_{i}^{g})||_{\vj} \neq 0$,
 since $|\rho_{g}^{\vj}(z_{i})|$ is always strictly positive.
 Let us consider such a point $(z_{i})\in P_{\vj}$. From property 1 we infer that if $(z_{i})$ is a critical point of 
 $||N(z_{i})||_{\vj}$, then it necessarily lies on the constraint surface $P^{(0)}_{\vj}$.
 Moreover, since the second derivative is always negative, 
 this means that this critical point is a maximum. This proves property 2.

 \subsection{Back to the Guillemin--Sternberg isomorphism}

 We can in fact show more: 
 consider a point $(z_{i})\in P_{\vj}$ with nonzero norm $||N(z_{i})||_{\vj} \neq 0$, and 
 let us
 look at the closure of the orbit $O_{z_{i}}= \mathrm{SL(2,\mathbb{C})}\cdot (z_{i})$.
 Since this is a closed submanifold of a compact space and since we established that 
 $||N(z_{i}^{g})||_{\vj}$ is a concave function of $g$,
  it reaches its maximum for a given element of the closure of 
 ${O}_{z_{i}}$. Because of property 1
 this maximum occurs in the constraint space $P^{(0)}_{\vj}$.
 The orbit space $ \mathrm{SL(2},\mathbb{C})\cdot P^{(0)}_{\vj}$ is an open neighborhood of $P^{(0)}_{\vj}$, because 
 at a given point $(z_{i}) \in  P^{(0)}_{\vj}$ the vectors $J(\widehat{X})$ are all orthogonal to tangent vectors 
 in  $ P^{(0)}_{\vj}$.
 The latter follows from
 \be\nonumber
 G(\xi,J(\widehat{X}))(z_{i})  = \xi H_{X}(z_{i}) =0\,,\quad \mathrm{when}\quad (z_{i}) \in P^{(0)}_{\vj}\,,\quad \xi \in T_{(z_{i})}P^{(0)}_{\vj}.
 \ee
Since  $ \mathrm{SL(2},\mathbb{C})\cdot P^{(0)}_{\vj}$ is an open neighborhood of $P^{(0)}_{\vj}$, the orbit $O_{z_{i}}$ intersects it.
Therefore, $z_{i} \in  \mathrm{SL(2},\mathbb{C})\cdot P^{(0)}_{\vj}$.
 
This proves the fundamental result of  Guillemin--Sternberg: namely, the characterisation of the orbit space 
$$ P^{s}_{\vj} \equiv \mathrm{SL(2},\mathbb{C})\cdot P^{(0)}_{\vj}=\{ z_{i}\in P_{\vj}\;|\;  ||N(z_{i})||_{\vj} \neq 0 \}$$ 
as the space, where the norm of the coherent intertwiner does not vanish\footnote{To be precise, Guillemin and Sternberg use a characterization in terms
of non--vanishing invariant holomorphic sections, but this is equivalent to the non--vanishing of coherent intertwiners, as we see below.}.

We can now finally tighten one of our loose ends and get the final characterisation of the stable set $P^{s}_{\vj}$.
First we expand the coherent state norm 
 in terms of the usual  
orthonormal intertwiner basis $|\vj, i\ket$ diagonalising the operator $(X_{1}+X_{2})$:
\be
\bra \vj, z|| \vj, z\ket = \sum_{i} \rd_{i} |S_{i}^{\vj}(z_{i})|^{2},\quad \mathrm{with}\quad S_{i}^{\vj}(z_{i})\equiv \bra \vj,i || \vj, n(z)\ket\,.
\ee
From this expansion one concludes that $||N(z_{i})||_{\vj} \neq 0$ iff there exists at least one holomorphic section
which does not vanish at $z_i$. Thus, the previous characterisation is equivalent to saying that $ P^{s}_{\vj} \equiv \mathrm{SL(2},\mathbb{C})\cdot P^{(0)}_{\vj}$
 is the space of points, where at least one invariant holomorphic section does not vanish.

We can now use the analysis  of the 4--valent sections performed in the previous section.
From (\ref{4-v}) one can infer that  if $\vj$ is regular 
(i.e.\ $j_{1}\pm j_{2}\pm j_{3}\pm j_{4}\neq 0 $ for any choice of sign) then for any nonzero section $S^{\vj}$,
we have that  $S^{\vj}(z_{i}) = 0$ when $Z(z_{i})= 0,1,\infty$.
On the other hand, it is clear that when $Z(z_{i})\neq 0,1,\infty$ (i.e.\ $z_{i}\neq z_{j}$) we can always find a section such that
$S^{\vj}(z_{i})\neq 0$.

More precisely, it is immediate from (\ref{4-v}) that $S^{\vj}(z_{i}) \neq 0$ if $z_{i}\neq z_{j}$ for all nonzero sections if $\vj$ is regular.
On the other hand, if for instance $j_{1}+j_{2}>j_{3}+j_{4}$, then $S^{\vj}(z_{i}) \to 0$ when $z_{12} \to 0$ (that is, when $Z\to \infty$).
Thus, what we have obtained is the characterisation of the orbit space as
$$ P^{s}_{\vj} \equiv \mathrm{SL(2},\mathbb{C})\cdot P^{(0)}_{\vj} = \mathbb{C} \backslash \{0,1,\infty\} \times  \mathrm{SL(2},\mathbb{C})$$
for regular\footnote{If $j_{i}$ is not regular, then it is possible to find a section which is nonzero even for $Z=0$ or $1$ or $\infty$,
so that the set $P^{s}$ is, in fact, bigger and includes also degenerate tetrahedra of zero volume.
For instance, in the case where $j_{1}+j_{2}=j_{3}+j_{4}$  there exists a section which is nonzero even when evaluated at $z_{12}=0$ or $z_{34}=0$.
} $j_{i}$.

\subsection{The main correspondence}

We now want to establish explicitly  the isomorphism between the Hilbert space $H_{\vj}$ which is the usual Hilbert space 
of SU(2) intertwiners and the Hilbert space $H_{\vj}^{(0)}$ obtained by first reducing and then quantizing. The latter is the span of 
coherent intertwiners satisfying the closure constraint.
We impose the restriction that the set  $j_{i}$ is regular, i.e.\  $j_{1}\pm j_{2}\pm j_{3}\pm j_{4}\neq 0 $ for any choice of sign.
We start from the decomposition of the identity of $H_{\vj}$, given by 
 \be 
 \mathbbm{1}_{\vj}= \prod_{i}\rd_{j_{i}} \int_{P_{\vj}} \prod_{i} \rd^{2} N(z_{i})\, ||\vj,N(z_{i})\ket \bra \vj, N(z_{i})||\,.
  \ee
The integral over $P_{\vj}$ can be replaced by the integral of the stable set $P_{\vj}^{s}$, since the complement is of measure zero.
  This integral can be further expanded as an integral over the constraint space and the $\mathrm{SL(2},\mathbb{C})$ orbit space
  due to the isomorphism 
  $$P_{\vj}^{s}= P_{\vj}^{(0)}/{\mathrm{SU(2)}} \times \mathrm{SL(2},\mathbb{C})=
 P_{\vj}^{(0)} \times \mathrm{SL(2},\mathbb{C})/{\mathrm{SU(2)}} $$  
 proven in the previous section.
 In the following, we will view the space $  \mathrm{SL(2},\mathbb{C})/{\mathrm{SU(2)}}\equiv H_{3}$ 
 as the space\footnote{The isomorphism between $\mathrm{SL(2},\mathbb{C})/\mathrm{SU(2)}$ and 
 $H_{3}$ is given by $g\to h^{2}=gg^{\dagger} $.
 This space is also the 3--dimensional Hyperbolic space.}  of positive hermitian two by two matrices of determinant $1$---that is, as a subspace of $\mathrm{SL(2},\mathbb{C})$.
 This space has a unique  invariant measure and we normalize this measure to be 
 $$\d e^{iX} =\left(\frac{2}{|X|} \sinh\frac{|X|}{2}\right)^{2} \prod_{a} \d X^{a}\quad \mathrm{with} \quad X= X^{a}\tau_{a}\in \mathfrak{su(2)}\,,\quad |X|^{2}=X^{a}X_{a}\,. $$
 
  Taking advantage of this decomposition we can now  show that 
  \be\label{int}
  \int_{P^{s}_{\vj}} \prod_{i} \rd^{2} N_{i} \,\, f(N_{i})= \int_{P_{\vj}^{(0)}} \rd\mu^{(0)}(N_{i}) \mathrm{det}\left(G(N_{i})\right)
   \left[\int_{H_{3}} \prod_{i}|\rho_{h}(N_{i})|^{2} f(h\cdot N_{i}) \rd h  \right]\,,
  \ee 
  where $\rd \mu^{(0)}(N_{i})$  is the SU(2) invariant measure on $P_{\vj}^{(0)}$: 
  $$\rd \mu^{(0)}(N_{i}) \equiv  \prod_{i} \rd^{2} N_{i}\; \delta^{(3)}\left(\sum_{i} j_{i} N_{i}\right),$$
  and $\mathrm{det}\left(G(N_{i})\right)$ is the determinant of the metric  (\ref{Gdef}) which has been computed in \cite{LivineS}:
  \bea
  \mathrm{det}\left(G(N_{i})\right) &=& \left(\sum_{i}j_{i}\right) \sum_{i>j} j_{i}j_{j} |[N_{i},N_{j}]|^{2} - \frac16 \sum_{i,j,k} j_{i}j_{j}j_{k} ([N_{i},N_{j}]|N_{k})^{2}\\
  &=& \frac{\left(\sum_{i}j_{i} \right)V^{2}(L) }{j_{1}j_{2}j_{3}j_{4}} \left(\sum_{i>j} j_{i}j_{j}|L^{ij}|^{2} - V^{2}(L)\right)\,.
  \eea
The second equality is only valid if the $N_{i}$'s satisfy the closure constraint.
In  this equality  we expressed the determinant in terms of the edge lengths $L^{ij}$ (see eq.\ \eq{edgelength})  and 
    the volume $V(L)$ of the tetrahedron\footnote{More precisely, $V(L) \equiv \mathrm{det}(L^{14},L^{24},L^{34})$  
    is 6 times the volume of the tetrahedron.} using the relations established in section \ref{tet}.
    It is interesting to note that the degenerate configurations are supressed by the presence of this determinant.

    In order to derive this integration formula we first recall that $G(X,Y)(N_{i}) = J(\widehat{X}) H_{Y} (N_{i})$. Thus, we have the identity
    \be
   1=  \mathrm{det}(G(N_{i})) \int_{H_{3}} \d h\;  \delta^{(3)}\left( h\cdot H \right)= \mathrm{det}(G(N_{i})) \int_{H_{3}} \d h\; \delta^{(3)}\left( h\cdot \left(\sum_{i} j_{i} N_{i}\right) \right)\,.
    \ee
    Inserting this identity in the LHS of (\ref{int}) and performing a change of variables $N_{i}\to h\cdot N_{i}$, one gets the RHS, since 
 \bea
 \rd^{2} (h\cdot N(z)) = |\rho_{h}(z)|^{2}  \rd^{2}  N(z)\,.
 \eea
     This proof is similar to the Faddeev--Popov proof of gauge fixing.

Given this lemma and the invariance property (\ref{stateinv}) of the state one arrives at
\be 
\label{mainformula}
\mathbbm{1}_{\vj}= \prod_{i}\rd_{j_{i}} \int_{P_{\vj}^{(0)}} \rd \mu_{\vj}(N_{i})\;  ||\vj,N_{i}\ket \bra \vj, N_{i}||\,,
\ee
where   the ``quantum'' integration measure $\rd \mu(N_{i})$ differs from the ``classical'' integration measure $\mu^{(0)}(N_{i})$
by a factor which is SL(2,$\mathbb{C}$) invariant and dependent on $\vj$:
  \be\label{meas}
  \rd \mu_{\vj}(N_{i}) \equiv \rd \mu^{(0)}(N_{i}) \mathrm{det}(G(N_{i})) 
   \left[\int_{H_{3}} \prod_{i}|\rho_{h}(z_{i})|^{2(j_{i}+1)} \rd h  \right]
   \ee
 What is remarkable about this identity is the fact that the integration is only over the 
constraint surface or the set of tetrahedra.
  This shows that coherent states which satisfy the closure constraint
  provide a basis of $H_{\vj} =\left(V^{j_{1}}\cdots \otimes V^{j_{4}}\right)^{\mathrm{SU(2)}}$. 
Since their labels correspond to classical tetrahedra, this means that these states have a very clear 
 geometrical interpretation as tetrahedra with fixed areas and no longer need any interpretation as fuzzy tetrahedra, as was the case in the works of 
 Barbieri \cite{Barbieri} and Livine--Speziale \cite{LivineS}.
 

 We finally want to establish the asymptotic property of the measure term when the spins are uniformely rescaled, i.e.\ $j_{i} \to \lambda j_{i}$.
 The measure (\ref{meas}) involves an integral of a weight which is always smaller than one.
 In the large $\lambda$ limit the integral therefore localises on its maximum point \cite{Ho1}. Since the $N_{i}$'s meet the constraint, 
 the maximum is at $h=1$. At this point the value of the integrand is equal to one and the integral is just determined  asymptotically 
 by the determinant of the matrix of quadratic fluctuation around the maximum. As we have established in eq.\ (\ref{G}), this matrix is given by $G$,
 that is, $\sum_{i} 2j_{i} \ln |\rho_{e^{iX}}(z_{i})| = -  \lambda G(X,X) + O(X^{3})$, therefore
 \be
  \rd \mu_{\vj}(N_{i}) \sim \rd \mu^{(0)}(N_{i})   \left(\frac{\pi}{\lambda}\right)^{\frac32} \frac{1}{\sqrt{\mathrm{det}(G(N_{i}))}}\,.
  \ee

\section{FK model and asymptotics for a single 4--simplex}

\subsection{FK$\gamma$ model}

In this section, we will apply our central result, eq.\ \eq{int}, to the FK$\gamma$ spin foam model \cite{FK}.
For $\gamma < 1$, this model is identical to the EPRL model \cite{ELPR}.

The model is defined on a two dimensional ``spine'' $\cal{S}$ which is the 2--dimensional skeleton of the 
intersection of $\Delta^{*}$ with $\Delta$, where $\Delta$ is a simplicial complex
 (for more details, see ref.\ \cite{FK} and \cite{CF1}). We denote by  $v$, $e$ and $f$ the edges vertices, edges and faces of the dual complex $\Delta^*$.
 We assign SU(2) $\times$ SU(2) group elements $\gb_{ev}=(g_{ev}^{+},g_{ev}^{-})$ to each oriented pair $ev$, with the convention that $g_{ve}= g_{ev}^{-1}$, 
  a point  of $S^{2}$  denoted $N_{ef}$ to every 
 pair $ef$, and a pair of spins $j^{\pm}=|\gamma^{\pm}|j_{f}$ to every face $f$\footnote{$\gamma^{\pm}$  are determined by the Immirzi parameter $\gamma$ 
as the smallest integers such that \cite{CF2}
\be
\frac{\gamma_{+}}{\gamma_{-}}=\frac{\gamma+1}{\gamma-1}\,,\quad \gamma^{+}>0\,.
\ee 
}.  
 A pair $(vf)$ is called a wedge and it is the meeting point of two edges $e,e'$.
The partition function of the model is given by
\bea
{Z = \sum_{j_f} \int \prod_{f, e\subset f} \d_{j^+_f}\d_{j^-_f}\d N_{ef} 
 \prod_{v, e\supset v} \d \gb_{ev}\; \prod_{v, f\supset v}A_{vf} (j_{f}^{\pm}, \gb_{ev},N_{ef})}\,,  
 \eea
where for $\gamma < 1$ the amplitude of the wedge $vf$ is\footnote{The corresponding amplitude for the case $\gamma > 1$ would be
\be
 A_{vf} (j_{f}^{\pm}, \gb_{ev},N_{ef}) = 
\b j^+_f, N_{ef} | T^{j^+_f}\!\!\left(g^+_{ev} g^+_{ve'}\right) | j^+_f, N_{e'\!f}\ket
\overline{\b j^-_f, N_{ef} | T^{j^-_f}\!\!\left(g^-_{ev} g^-_{ve'}\right) | j^-_f, N_{e'\!f}\ket}\,,
\ee
with the overline denoting complex conjugation.
 For sake of brevity, we only treat the case $\gamma < 1$ here. The derivation and results for $\gamma > 1$ are essentially the same
up to minor technical modifications.}
 \be
 A_{vf} (j_{f}^{\pm}, \gb_{ev},N_{ef}) = 
\b j^+_f, N_{ef} | T^{j^+_f}\!\!\left(g^+_{ev} g^+_{ve'}\right) | j^+_f, N_{e'\!f}\ket
\b j^-_f, N_{ef} | T^{j^-_f}\!\!\left(g^-_{ev} g^-_{ve'}\right) | j^-_f, N_{e'\!f}\ket\,.
\ee

This amplitude can be described as a contraction of tensor products of coherent intertwiners from each edge: that is, if one denotes $\vj_{e} \equiv (j_{f})_{f\subset e}$ and $ \vec{N}_{e} \equiv (N_{ef})_{f\subset e}$, one can obtain $Z$ by contracting the invariant tensor
\be
\bigotimes_{e} \int \prod_{f\supset e} \d N_{ef} \left( ||\vj^{\,\,+}_{e},\vec{N}_{e}\ket \bra \vj^{\,\,+}_{e},\vec{N}_{e}|| \otimes  ||\vj^{\,\,-}_{e},\vec{N}_{e}\ket \bra\vj^{\,\,-}_{e},\vec{N}_{e}||\right)\,.
\ee 
When viewed in this way, it becomes clear that we can apply the main result proven in the previous section: 
first one replaces the integral over unconstrained $N_{ef}$ by an integral over $P^{(0)}_{\vj} \times H_{3}$,
 and then one uses the $\mathrm{SL(2,}\mathbb{C})$ covariance in order to integrate out the dependence on the ``fiber'' $H_{3}$.
Thus, we get the amplitude
\bea
Z \label{Zdef}
 &=& \sum_{j_f} \prod_e  \int\limits_{P^{(0)}_{\vj}} \!\!\d{\mu}^{\vj_e}_e(N_{ef}) 
\int \prod_{v, e\supset v} \d \gb_{ev} \; \prod_{v, f\supset v}A_{vf} (j_{f}^{\pm}, \gb_{ev},N_{ef})\,,
\eea
where the new measure for $N_{ef}$, $f\supset e$, is 
$$
\d{\mu}_e^{\vj_e}(N_{ef}) \equiv \left(\prod_{f\supset e} \d_{j^+_f}\d_{j^-_f}
\d^{2} N_{ef}\right)\,  \delta^{(3)}\left(\sum_{f\supset e} j_f N_{ef}\right) \mathrm{det}(G_{\vj_e}(N_{ef}))
 \left[\int_{H_{3}} \prod_{f\supset e}|\rho_{h}(z_{ef})|^{2(j_f^{+} +j^{-}_f+1)} \rd h  \right]
\,.
$$
The key point is that now we only integrate over vectors  that satisfy the closure constraint, i.e.\ $\sum\limits_{f\supset e} j_f N_{ef}=0$ for each edge $e$.
 
So far we have only described the amplitude for a closed 2d spine $\cal S$,
but it is straightforward to generalize it to an amplitude with boundary states.
Such an extension is needed in the computation of the graviton propagator (see for example \cite{Rov1,Alesci1, Alesci2}).
 
It is known that in the case $\gamma < 1$ the boundary states are SU(2) spin networks \cite{EPR2,CF2}.
Such states are characterised by a choice of a 4--valent graph $\Gamma$, which is the boundary 
 of the spine $\cal{S}$, a choice of SU(2) representations $k_{\bar{e}}$ associated with 
 the edges $\bar{e}$ of $\Gamma$ and the choice of an intertwiner $i_{\bar{v}}\in \left(\otimes_{\bar{e} \supset \bar{v}} V_{k_{\bar{e}}}\right)^{\mathrm{SU(2)}}$  for each vertex $\bar{v}$ of $\Gamma$.
 $i_{\bar{v}}$ is a label representing either a usual spin network intertwiner (in which case it is a half--integer) or 
 a coherent intertwiner (in which case it is given by four unit vectors).
 The edge label $k_{\bar{e}}$ of $\Gamma$ is related to the face label $j_f$ of a face intersecting the boundary: namely, $k_{\bar{e}} = j^+_f + j^-_f$ or $j_f = k_{\bar{e}} / (\gamma^+ - \gamma^-)$ (since $\gamma^-<0$ for $\gamma<1$).

In order to define the amplitude we just have to distinguish between the bulk edges  $e$
of the spine and the boundary edges of the spine which are in one--to--one correspondence with 
boundary vertices $\bar{v}$.
Similarly, we distinguish the bulk faces $f$ from boundary faces which are in one--to--one correspondence 
with boundary edges $\bar{e}$.
Given these boundary data, we can specify the boundary amplitude as
\be
Z(\Gamma, k_{\bar{e}},i_{\bar{v}}) =  
\int  \prod_{\bar{v}, \bar{e}\supset\bar{v}} \d_{k_{\bar{e}}} \d N_{\bar{v}\bar{e}}
\;
Z(j_{\bar{e}}, N_{\bar{v}\bar{e}})
\prod_{\bar{v}} \bra \vec{k}_{\bar{v}}, \vec{N}_{\bar{v}} \| \vec{k}_{\bar{v}}, i_{\bar{v}}\ket\,.
\label{boundaryamplitude}
\ee
Here, $j_{\bar{e}} = k_{\bar{e}} / (\gamma^+ - \gamma^-)$, and
$Z(j_{\bar{e}}, N_{\bar{v}\bar{e}})$ is defined like the partition function \eq{Zdef} except that the integration and summation
extends only over bulk variables:
\be
Z(j_{\bar{e}}, N_{\bar{v}\bar{e}})
= 
\sum_{j_f} \int \prod_{f, e\subset f} \d_{j^+_f}\d_{j^-_f}\d N_{ef} 
\prod_{v, e\supset v} \d \gb_{ev}\;\prod_{v, f\supset v}A_{vf} (j_f^{\pm}, \gb_{ev},N_{ef})\,.  
\ee
One can show that this definition of the boundary amplitude agrees with the one given in \cite{CF1}\footnote{For this one uses the coherence property of coherent states, that is, the identity
\be
|j^{+}, N\ket \otimes |j^{-}, N\ket = |k, N\ket\quad \mathrm{for}\quad k= j^{+}+ j^{-}.
\ee}.
As before, we can apply the result of the previous section and replace the integral in \eq{boundaryamplitude} by an integral with measure $\mu$
where the closure constraint is imposed strongly:
\be
Z(\Gamma, k_{\bar{e}},i_{\bar{v}}) = 
\prod_{\bar{v}}\int\limits_{P^{(0)}_{\vj}} \!\! \d{\mu}_{\bar{v}}^{\vj_{\bar{v}}}(\vec{N}_{\bar{v}})
\;
Z(j_{\bar{e}}, N_{\bar{v}\bar{e}})
\prod_{\bar{v}} \bra \vec{k}_{\bar{v}}, \vec{N}_{\bar{v}} \| \vec{k}_{\bar{v}}, i_{\bar{v}}\ket
\label{boundaryamplitudewithnewmeasure}
\ee
with
\be
\d{\mu}_{\bar{v}}^{\vj_{\bar{v}}}(\vec{N}_{\bar{v}}) \equiv 
\left(\prod_{\bar{v}, \bar{e}\supset\bar{v}} \d_{k_{\bar{e}}} 
\d^{2} N_{\bar{v}\bar{e}}\right)\,  
\delta^{(3)}\left(\sum_{\bar{e}\supset \bar{v}}k_{\bar{e}} N_{\bar{v}\bar{e}}\right) 
\mathrm{det}(G_{\vj_{\bar{v}}}(N_{\bar{v}\bar{e}}))
\left[\int_{H_{3}} \prod_{\bar{e}\supset \bar{v}} |\rho_{h}(z_{\bar{v}\bar{e}})|^{2(k_{\bar{e}}+1)} \rd h  \right]
\,
\ee
and $k_{\bar{e}}=j_{\bar{e}}^{+} +j^{-}_{\bar{e}}.$

\subsection{Asymptotics of vertex amplitude}
\label{asymptoticsofvertexamplitude}

We now want to illustrate that the techniques used in \cite{CF2} can be naturally extended to analyse the asymptotics
of boundary amplitudes. Here, we will focus on the case of one 4--simplex (i.e.\ the vertex amplitude) reserving the general study for the future.

\psfrag{v}{$v$}
\psfrag{f}{$f_{12}$}
\psfrag{g1}{$g_{e_1v}$}
\psfrag{g2}{$g_{e_2v}$}
\psfrag{n1}{$n_{e_1f_{12}}$}
\psfrag{n2}{$n_{e_2f_{12}}$}
\psfrag{u1}{$u_{e_1}$}
\psfrag{u2}{$u_{e_2}$}
\psfrag{e1}{$e_1$}
\psfrag{e2}{$e_2$}
\pic{4simplex}{Dual of a 4--simplex with boundary: the boundary of the 4--simplex intersects dual edges at their center. 
The intersection points provide the base point for boundary data $n_{e_if_{ij}}, u_{e_i}$.}{4.5cm}{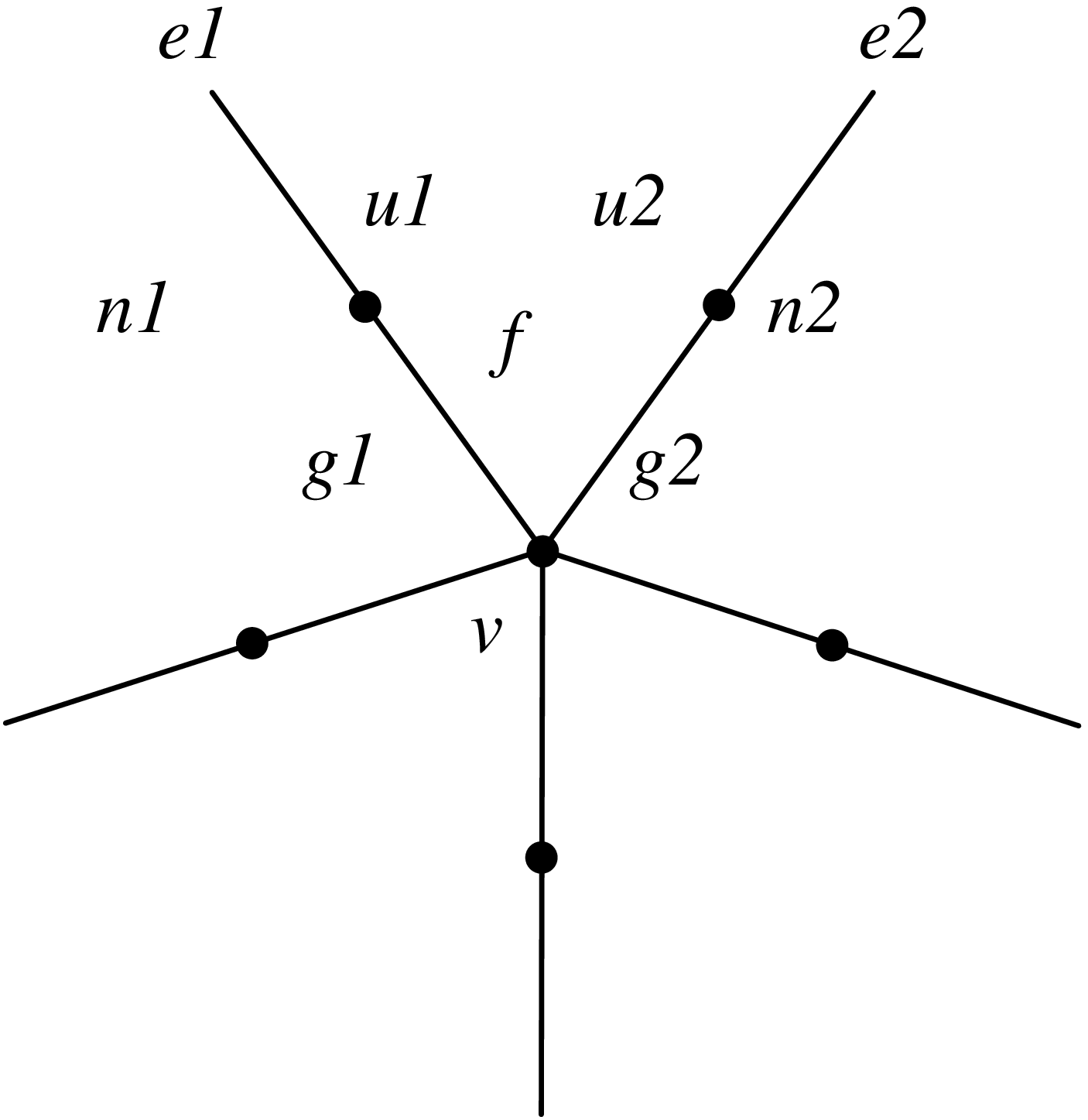}

To describe the contribution of a single 4--simplex dual to a vertex $v$ we number the edges at the vertex $e_i$, $i=1,\ldots,5$ (see \fig{4simplex}).
Moreover, it will be convenient 
to abbreviate the notation and set $\gb_i \equiv \gb_{ve_i}$, $n_{ij} \equiv n_{e_if_{ij}}$, $u_i \equiv u_{e_i}$, $j^\pm_{ij} \equiv j^\pm_{f_{ij}}$ etc.
Then,
\be
A_v(j_{ij}^{\pm},N_{ij}) = \int \prod_i \d g^+_i \d g^-_i
\prod_{i < j} 
A_{ij}(j^{\pm}_{ij}, \gb_{i}, N_{ij})
\label{vertexamplitude}
\ee
where
\be
A_{ij}(j^{\pm}_{ij}, \gb_{i}, N_{ij}) =
\b j^+_{ij}, N_{ij} | T^{j^+_{ij}}\!\!\left(\left(g^+_i\right)^{-1} g^+_j\right) | j^+_{ij}, N_{ji}\ket
\b j^-_{ij}, N_{ij} | T^{j^-_{ij}}\!\!\left(\left(g^-_i\right)^{-1} g^-_j\right) | j^-_{ij}, N_{ji}\ket\,.
\ee
We assume that the closure constraint is imposed, since, as shown in the previous section, we only
have to integrate over constrained configurations in order to define the boundary state amplitude. Therefore, 
\be 
\sum_{j:j\neq i} j_{ij} N_{ij} =0\,.\label{clos}
\ee
Also, in order to use the results of \cite{CF2} we need to assume that the integral is only over 
nondegenerate configurations, i.e.\ we impose in the integral the restriction 
\be 
\det (g_{1} \hat{U},\cdots, g_{4} \hat{U}) \neq 0\,,
\label{nondegeneracycondition}
\ee
where $\Uh = (1,0,0,0)^T$ is a reference vector. We call the restricted nondegenerate amplitude $ A_{v}^{\mathrm{nd}}$.

Before we can state our result on the asymptotics, we need to introduce a number of definitions. Recall from refs.\ \cite{CF1,CF2} that the variables $j^\pm_f$ and $N_{ef}$ define
a bivector $X^\gamma_{ef}$ that represents the discrete analogue of the $B$--field. Its selfdual and anti--selfdual components are
\be
\Xb^\gamma_{ef} = (j^+_f N_{ef}, j^-_f N_{ef})\,,\qquad N_{ef} \equiv N^i_{ef}\sigma_i\,.
\ee
It is convenient to introduce a \textit{simple} bivector $X_{ef}$ by setting
\be
X^{\gamma\pm} \equiv \gamma^\pm X^\pm\,,
\qquad \mbox{so that}\qquad
\Xb_{ef} = (j_f N_{ef}, j_f N_{ef})\,,
\ee
and
\be
X^\gamma_{ef} = \frac{1}{2}\left(\gamma^+ + \gamma^-\right)\left(X_{ef} + \frac{1}{\gamma} \star X_{ef}\right)\,.
\ee
From lemma C.1.\ in \cite{CF1} we also know that this corresponds to a bivector
\be
X_{ef} = j_f\Uh\wedge \hat{N}_{ef}\,,
\ee
where $\Uh = (1,0,0,0)^T$ and $\hat{N}_{ef}$ is the \textit{4--vector} $\hat{N}_{ef} = (0,N_{ef})$ and $N_{ef}$ the previously defined 3--vector.
In terms of our index notation, this means that 
\be
X_{ij} = j_{ij} \Uh\wedge \hat{N}_{ij}\,.
\ee

Let us next define vectors
\be
V_{ij} = R(n_{ij},n_{ij}) (0,1,0,0)^T\,,\qquad W_{ij} = R(n_{ij},n_{ij}) (0,0,1,0)^T\,,
\ee
where $R(g^+,g^-)$ denotes the projection of the SU(2)$\times$SU(2) element $\gb = (g^+,g^-)$ to SO(4). 
We will see in appendix \ref{derivationofasymtptotics} that $V_{ij}$ and $W_{ij}$ are vectors in the plane orthogonal to $X_{ij}$.
Given the vectors $\Uh$, $\hat{N}_{ij}$, $V_{ij}$ and $W_{ij}$, we specify angles $\theta_{ij}$ and $\thetat_{ij}$ by
\bea
\cos\theta_{ij} = \Uh\cdot g_{ij}\Uh\,, &\qquad&
\sin\theta_{ij} = \Uh\cdot g_{ij}\hat{N}_{ji}\,,
\label{defU} \\
\cos\thetat_{ij} = V_{ij}\cdot g_{ij}V_{ji}\,, &\qquad&
\sin\thetat_{ij} = V_{ij}\cdot g_{ij} W_{ji}\,,
\label{definitionVW}
\eea
where $g\Uh$, $g\hat{N}$ etc.\ stands for the action of the SO(4) element $g$ on 4--vectors.
These angles are needed to express the asymptotics of the vertex amplitude.

The asymptotics is the result of a saddle point approximation of the integral over $\gb_i$ in \eq{vertexamplitude}.
Since 
\be
\left|\b j^+ N_{i} | g_{i}^{-1} g_j  | j^+ N_{j}\ket\right| = \left|\frac12\left(1 + \left(g_{i}N_{i}g_{i}^{-1} | g_{j}N_{j}g_{j}^{-1}\right)\right)\right|^{j^+}\leq 1\,,
\ee
with the equality satisfied iff $g_{i}N_{i}g_{i}^{-1}=g_{j}N_{j}g_{j}^{-1}$, it follows that 
the saddle points (the points were the integrand is maximal) are determined by the equations
\be 
\label{staphase}
g_iX_{ij}g_i^{-1} =  g_j X_{ji} g_j^{-1} \,.
\ee
This equation should be supplemented by  the closure constraint (\ref{clos}) of  the boundary data, which can be written as 
\be
\sum_{j\neq i} X_{ij} = 0\,.
\ee

When a solution exists and nondegeneracy is assumed, these equations imply that there is a 4--simplex at the vertex $v$
which is consistent with the tetrahedra in the boundary\footnote{See \cite{BJ} for a canonical analysis of the same equations.}. 
This follows from a straightforward extension of proposition\ VII.1, \cite{CF2} to a complex with boundary. The difference to the analysis in \cite{CF2} is only that the boundary conditions for the 4--simplex are now fixed by the boundary tetrahedra, and not by neighbouring 4--simplices.

The geometry of this 4--simplex is determined by the 4--simplex tetrad $U_{i}$ (see def.\ VI.3 and prop.\ VI.4 of \cite{CF2}), i.e.\ a set of five 4d vectors which satisfy the closure condition $\sum_{i} U_{i}=0$  and whose direction is, up to sign, given by $g_{i}\hat{U}$:
\be
\frac{U_{i}}{|U_{i}|} = \pm g_{i}\hat{U}\,.
\label{UUi}
\ee
The bivectors $X_{ij}$ arise from wedge products of the tetrad vectors:
\be
g_iX_{ij}g_i^{-1} = \epsilon V_4(v) ( U_i \wedge U_j )\,.
\ee
$\epsilon$ is a sign factor and determined by the boundary data. The group elements $g_i$ are, up to sign, equal
to the spin connection, which maps the vertex geometry of the tetrahedra determined by $U_{i}$ to the boundary geometry of the tetrahedra determined by $N_{ij}$.
More precisely, given the $U_i$, we can compute the geometry of any of the five tetrahedra of the 4--simplex. As a result of the stationary phase equation \eq{staphase}, this geometry is the same as the one derived directly from the boundary data (see \sec{tet}). They only differ by a rotation of frame specified by the spin connection. These results follow from proposition VII.1 and proposition B.2 of \cite{CF2}.

From equality (\ref{UUi}) and definition (\ref{defU}) one sees that the angles $\theta_{ij}$ are  the dihedral angles of the 4--simplex geometry determined by $U_{i}$ modulo $\pi$. The subtlety is that  $g_i\Uh$ can differ from the normalised tetrad  either by a sign or by a global reflection\footnote{This last point was initially overlooked
and corrected thanks to reference \cite{BarrettF}.} around $\hat{U}$, and thus $\theta_{ij}$ equals the dihedral angle only up to $\pi$ or up to a global sign.
The reflection around $\Uh$ does not affect the $\tilde{\theta}$ angles, since they correspond to rotations  in a plane orthogonal to $\hat{U}$.

Here, we will restrict our analysis to the case where the spins $j_{ij}^{\pm}$ are integers. In this case, the difference between $\theta_{ij}$ and $\theta_{ij} + \pi$
is of no consequence in the asymptotics (see eq.\ \eq{asymptoticsAv} below).

Based on the saddle point approximation we can now deduce the asymptotics of the integral \eq{vertexamplitude}. When the eqns.\ \eq{staphase} have a solution, we find that
\be
A_v^{\mathrm{nd}} \sim \frac1{(4\pi)^{4}}\frac{1}{\sqrt{ \mathrm{det}(G_{+})\mathrm{det}(G_{-})}}
\exp\left[-\irm\sum_{i<j}
\left((\gamma^+ + \gamma^-) j_{ij} \theta_{ij} + (\gamma^+ - \gamma^-) j_{ij} \thetat_{ij} \right) \right] + (\theta\to -\theta)\,,
\label{asymptoticsAv}
\ee
where the second term is obtained by the reflection $(\theta,\tilde{\theta}) \to (-\theta,\tilde{\theta})$.
When there are no nondegenerate solutions, the amplitude is exponentially suppressed.
In the latter case, there is no nondegenerate 4--simplex which is consistent with the boundary tetrahedra.

Since  $\clA_{ij} = (\gamma^+ + \gamma^-) j_{ij} / 2$ is the area of the triangle dual to the face $ij$ and 
$\theta_{ij}$ is the dihedral angle, the first term in the exponential is the 4d Regge action of the 4--simplex.
The second term depends on the additional angles $\thetat_{ij}$. One can show (directly or using the results of \cite{CF2}) that these terms drop out when the saddle point approximation is applied to a bulk region consisting of several 4--simplices. More precisely, for a face $f$ in the bulk $\sum_{f\supset e}\tilde{\theta}_{ef} =0$ modulo $\pi$. Therefore, eq.\ \eq{asymptoticsAv} is consistent with the bulk asymptotics derived in our previous paper \cite{CF2}.
It can be also understood that the extra angle $\thetat_{ij}$ is essentially due to a U(1) gauge symmetry of coherent states.
Indeed, we have labelled coherent states so far by elements $N\in \mathrm{SU(2)}/\mathrm{U(1)}$ and corresponding group elements $n\in \mathrm{SU(2)}$. 
We could label them just as well by SU(2) elements $n' = n\, \e^{\irm\frac{\phi}{2}\sigma_{3}}$. If we do so, we see that the vertex amplitude becomes\footnote{Observe that for  $\gamma<1$ we have that $\gamma^{-}<0$.}
\be 
A_{v}(n'_{ij}) = A_{v}(n_{ij})\, \e^{\irm (\gamma^+ - \gamma^-)j_{ij} \sum_{i<j} \left(\phi_{ij} - \phi_{ji}\right)}\,.
\ee
Thus, the angles $\tilde{\theta}_{ij}$ can be locally gauged to zero.

The determinants $\mathrm{det}(G_{\pm})$ are determinants of $4\times 3$ by $4\times 3$ matrices: the rows and columns are labelled by pairs $ai$, where $i$ denotes four tetrahedra of the 4-simplex, and $a$ is a 3--dimensional index labelling a basis of $\mathfrak{su(2)}$. The matrix elements of $G_{\pm}$ are sums of the hermitian matrices (\ref{h}):
\bea
G_{\pm}^{ia,ib} &=&  \sum_{j:j\neq i }  {j_{ij}^{\pm}} \left(\delta^{ab} - N_{\pm ij}^{a}N_{\pm ij}^{b} \right)\,, \\
G_{\pm}^{ia,jb} &=& -{j_{ij}^{\pm}} \left(\delta^{ab} - N_{\pm ij}^{a}N_{\pm ij}^{b}  - i \epsilon^{ab}{}_{c} N^{c}_{\pm ij}(z)\right)\,,\qquad i\neq j\,,
\eea
with $N_{\pm ij} \equiv g^{\pm}_{i}N_{ij}(g^{\pm}_{i})^{-1}$.
A detailed derivation of formula \eq{asymptoticsAv} is provided in appendix \ref{derivationofasymtptotics}.

Eq.\ \eq{asymptoticsAv} describes the asymptotics of the vertex amplitude $A_v^{\mathrm{nd}}$, in which the integration over $g_i$ is constrained  by the nondegeneracy condition \eq{nondegeneracycondition}. Therefore, there can only appear saddle points corresponding to nondegenerate 4--simplices. For these points, our formula agrees, up to numerical factors, with the asymptotics obtained in ref.\ \cite{BarrettF}. In the latter work the vertex amplitude is treated without any restriction, so there appear additional terms corresponding to degenerate configurations.

\section{Summary and discussion}

Let us summarize the contents and results of this paper: 
we followed a reduction--before--quantization approach to construct the Hilbert space of a tetrahedron in $\bR^3$.
Guided by theorems of Guillemin \& Sternberg and Hall \cite{GS,Hall} we related this quantization to the more conventional Dirac quantization, where one quantizes first and imposes the constraints afterwards. We started by describing the classical phase space of tetrahedra and the Guillemin--Sternberg theorem at the classical level (relating the phase spaces $P^{(0)}/{\mathrm{SU(2)}}$ and $P^{s}/{\mathrm{SL(2,}}\mathbb{C})$). 
Then, we reviewed properties of group coherent states which we need to connect classical phase space and quantum theory.
Section \ref{QuantRed} outlined the content of the Guillemin--Sternberg isomorphism at the quantum level. 
We then gave a proof and explicit construction of this map for the quantum tetrahedron (\sec{Main}): 
the central formula is eq.\  \eq{mainformula}, which expresses the resolution of identity by an integral over
coherent states labelled by tetrahedra. The underlying idea is the following: coherent states provide an overcomplete set of states and they are labelled by phase space variables. By Guillemin \& Sternberg's theorem the unconstrained phase space is naturally fibered over the constraint phase space with the fiber being 
isomorphic to the imaginary  part of the gauge group.
 Therefore, we can express the integral   over coherent states as an integral over the constrained variables and an integral over the hermitian elements of $\mathrm{SL(2,\bC)}$. Due to the $\mathrm{SL(2,\bC)}$ invariance of the state the fiber integral decouples via a Faddeev--Popov trick and gives an explicit measure factor.

In the last section, this result was applied to the FK$\gamma$ model in order to express the path integral in terms of tetrahedral variables\footnote{For $\gamma < 1$, this applies also to the EPRL model \cite{ELPR}, since in this case the FK and EPRL model are the same.}. Using this, we determined the asymptotics of the vertex amplitude in the nondegenerate sector. Up to numerical factors, our result coincides with the nondegenerate contribution in the asymptotic formula by Barrett et al.\ \cite{BarrettF}. 

A main motivation for this work came from a previous result, where we showed that in the semiclassical limit spin foam amplitudes reduce to the Regge action of discrete geometries \cite{CF2}. On the one hand, this showed a clear link between the spin foam model and discretized general relativity. On the other hand, it also made very clear that, a priori, spin foams are not geometries. The spin foam model can be seen as a first--order path integral, where the ``metric'' variables are not tetrads, but more general objects, that only become tetrads when we go on--shell w.r.t.\ the reality of the action and the equations of motion for the connection (which includes the closure constraint).

For this reason, it is important to know how the amplitudes behave as one moves away from the constraint surface. When computing a graviton propagator, for example, one has to integrate over \textit{all} fluctuations around a background, which includes those that do not fulfill the constraints.
Thus, we were interested in expressing the spin foam amplitude as a function of an on--shell background plus something else which would parametrize the points away from the constraint surface.

With equation \eq{mainformula} we have obtained precisely such a formula for the closure constraint. The configurations are labelled by geometrical tetrahedra and $\mathrm{SL(2,\bC)}$ elements: the former characterize the point on the constraint surface from which we start to generate an arbitrary point by application of  $\mathrm{SL(2,\bC)}$. Our formula tells us how the amplitude changes as a function of the $\mathrm{SL(2,\bC)}$ element, or equivalently, as a function of the distance from the constraint surface.  

This also answers another question that motivated our work: in \cite{CF2} an essential, but somewhat mysterious ingredient was the fact that the action possessed an imaginary part\footnote{By the action we mean here $S$ such that $A = \e^{\irm S}$, where $A$ is the amplitude.}. This imaginary part was absolutely essential in order to show that the amplitude had the right semiclassical limit: the condition that the imaginary part vanishes supplemented the equations of motion and ensured that only geometrical configurations dominate. We now see that this imaginary part is related to the existence of a natural metric on the Guillemin--Sternberg fibration of phase space. The latter measures how far we are from ``metricity''. Remarkably, as we have seen, some of the non--metrical dependence can be explicitly integrated out.

Our result is a further step in the direction initiated by Livine and Speziale's coherent state method: namely, the attempt to express quantum states in terms of geometrical quantities. This approach has already led to considerable progress and clarification in the construction of spin foam models.
With the new overcomplete set of states defined here, we go yet another step in this direction: now the states are not only labelled by spins and normal vectors, but, in addition, these normal vectors close, so they truly define a geometrical tetrahedron. In this sense, these states are closer to the notion of geometry than the spin network intertwiners, and one is more justified to call them states of quantum geometry.

We believe that the use of such states could have various advantages in spin foam models and also in canonical loop quantum gravity. 
It could give access to perturbation theory and Feynman diagrams, since the amplitude is now conveniently expressed as a function of the on--shell background and an ``off--shell parameter''. Thus, the path integral resembles a lattice path integral over a tetrad field, a connection and further parameters, and it seems conceivable that one can derive Feynman diagrams in a similar way as in lattice gauge theory.

The methods developed in our paper should also shed new light on the coherent states of canonical loop quantum gravity  \cite{AshtekaretalCoherentstatetransform}. A notable difference between our states and the ones described there is the fact that the latter states are designed to give semiclassical peakedness in both the connection and area, while our states are peaked on explicit tetrahedral geometries. It would be interesting to understand the relationship between the two approaches. 

The LQG coherent states arise in the definition of the coherent state transform 
$L^2(\clA) \to L^2_{\mathrm{hol}}(\clA^{\bC} )$ \cite{AshtekaretalCoherentstatetransform} and extend Hall's notion 
of group coherent states \cite{Hall}. In ref.\ \cite{ThiemannBahr}, Bahr and Thiemann have applied group averaging on these states, and obtained gauge--invariant coherent states which are the analog of the Livine--Speziale coherent states. Such states are not only invariant under the gauge group $\clG$, but also under the complexified group $\clG^{\bC}$, as in our case. What is missing so far in the analysis of these states (and this would constitute the analog of what we show here)
is the proof that the Hilbert space of gauge invariant states is in fact unitarily equivalent to the Hilbert space of holomorphic functionals 
$L^2_{\mathrm{hol}}(\clA^{s} / \clG^{\bC})$ with an appropriate measure, where $\clA^s$ is the analog of the semistable set $P^s_{\vj}$.
Such an isomorphism would imply that all gauge invariant coherent states are properly peaked and it would also provide an explicit parametrisation and scalar product on the space of coherent states on which the Gauss law is imposed. We hope to come back to these issues in the future by expanding on the techniques that we have developed here.

In summary: our work continues in the spirit of recent developments that emphasize the geometric aspect of spin foam models and loop quantum gravity.
We are hopeful that this will lead to further progress in linking the quantum with the classical theory and the spin foam with the canonical approach, and that it will bring us closer to answering the decisive question: namely, if this theory of quantum gravity is, in fact, consistent with classical general relativity.

\begin{acknowledgments}
We thank E. Alesci, E. Livine, B. Dittrich and S. Speziale for discussions. Research at Perimeter Institute is supported by the Government of Canada through Industry Canada and by the Province of Ontario through the Ministry of Research \& Innovation. 
\end{acknowledgments} 

\begin{appendix}

\section{Derivation of asymptotics}
\label{derivationofasymtptotics}

In this appendix we give more details on the derivation of the asymptotics \eq{asymptoticsAv}.
We will first focus on the edges $e_1$ and $e_2$ and write $N_1 \equiv N_{12}$, $N_2 \equiv N_{21}$ and $j \equiv j_{12}$ for simplicity.
The associated group elements are $n_1 \equiv n_{12}$ and $n_2 \equiv n_{21}$.

From the edges $e_1$, $e_2$ and the face $f_{12}$ we get the amplitude factor
\be
A_{12} = \b j^+ n_1 | (g^+_1)^{-1} g^+_2 | j^+ n_2\ket \b j^- n_1 | (g^-_1)^{-1} g^-_2 | j^- n_2\ket\,.
\ee
Let us set $n^\pm_2 = \left(g^\pm_1\right)^{-1} g^\pm_2 n_2$. Then,
\bea
\left|\b j^+ n_1 | \left(g^+_1\right)^{-1} g^+_2 | j^+ n_2\ket\right| &=& \left|\b j^+ N_1 | j^+ N^+_2\ket\right| = \left(\frac{1}{2}\left(1 + N_1\cdot N^+_2\right)\right)^{j^+}\,,
\label{modulusmatrixelement+} \\
\left|\b j^- n_1 | \left(g^-_1\right)^{-1} g^-_2 | j^- n_2\ket\right| &=& \left|\b j^- N_1 | j^+ N^-_2\ket\right| = \left(\frac{1}{2}\left(1 + N_1\cdot N^-_2\right)\right)^{j^-}\,,
\label{modulusmatrixelement-}
\eea
where $N^+_2$ and $N^-_2$ are the unit 3--vectors associated to the group elements $n^+_2$ and $n^-_2$.
On the other hand,
\bea
\b j^+ n_1 | \left(g^+_1\right)^{-1} g^+_2 | j^+ n_2\ket &=&  \left(\b {\sst\frac{1}{2}\,{-\frac{1}{2}}} | n^{-1}_1 \left(g^+_1\right)^{-1} g^+_2 n_2 | {\sst\frac{1}{2}\,{-\frac{1}{2}}}\ket\right)^{2j} = \e^{\irm j^+ \phi^+} \cos^{2j^+}\left(\varphi^+ / 2\right)\,, \\
\b j^- n_1 | \left(g^-_1\right)^{-1} g^-_2 | j^- n_2\ket &=&  \left(\b {\sst\frac{1}{2}\,{-\frac{1}{2}}} | n^{-1}_1 \left(g^-_1\right)^{-1} g^-_2 n_2 | {\sst\frac{1}{2}\,{-\frac{1}{2}}}\ket\right)^{2j} = \e^{\irm j^- \phi^-} \cos^{2j^-}\left(\varphi^- / 2\right)
\eea
for some angles $\phi^\pm$ and $\varphi^\pm$.

In the large spin limit, the moduli \eq{modulusmatrixelement+} and \eq{modulusmatrixelement-} are exponentially suppressed unless $N_1 = N^+_2 = N^-_2$. This condition is equivalent to the statement that
\be
X_{12} = g_1^{-1} g_2 X_{21} g_2^{-1} g_1\,.
\label{X1221}
\ee
If this is fulfilled, the moduli are 1, the angles $\varphi^\pm$ are zero, and there only remain the phases $\e^{\irm j^\pm\phi^\pm}$.
In this case,
\be
n^{-1}_1 \left(g^\pm_1\right)^{-1} g^\pm_2 n_2 = \e^{- \irm\phi^\pm J_3}
\label{g'12J3}
\ee
and
\be
A_{12} = \b j n_1 | g_1^{-1} g_2 | j n_2\ket \b j^- n_1 | (g^-_1)^{-1} g^-_2 | j^- n_2\ket 
= \e^{\irm \left(j^+ \phi^+ + j^- \phi^-\right)}\,.
\label{A12phase}
\ee
By the same token we obtain the equations
\be
X_{ij} = g_i^{-1} g_j X_{ji} g_j^{-1} g_i\,.
\label{Xijji}
\ee
for the other pairs $i, j$. 

As explained in \sec{asymptoticsofvertexamplitude}, if there is a solution to eqns.\ \eq{Xijji}, it implies 
the existence of a 4--simplex at $v$ whose area bivectors are the bivectors $\epsilon \star (g_i X_{ij} g_i^{-1})$. 
The vectors $g_i\Uh$ provide the normal vectors of tetrahedra at the basepoint $v$. Consequently, the angle $\theta_{12}$ defined by
\bea
\cos\theta_{12} &=& \Uh\cdot g_{12}\Uh\,, \\
\sin\theta_{12} &=& \Uh\cdot g_{12} \hat{N}_2\,,
\eea
is the dihedral angle\footnote{As we discussed in the main part, it is actually the dihedral angle modulo $\pi$, due to the signs in eq.\ \eq{UUi}.} $\theta_{12}$ between the tetrahedra corresponding to the edge $e_1$ and $e_2$.

To evaluate the phase in \eq{A12phase}, it is convenient to view the group elements in \eq{g'12J3} as the result of a gauge transformation which sends $n_1$ and $n_2$ to the identity:
\bea
g'^\pm_{12} &=& n^{-1}_1 \left(g^\pm_1\right)^{-1} g^\pm_2 n_2\,, \label{gaugeforn1} \\
n'_1 &=& \mathbbm{1}\,, \label{gaugeforn2} \\
n'_2 &=& \mathbbm{1}\,. \label{gaugeforn3}
\eea
By equation \eq{g'12J3} we have that $\left(g'_{12}\right)^\pm = \e^{- \irm\phi^\pm J_3}$.
The homomorphism $h: \mathrm{su(2)}\oplus\mathrm{su(2)}\to\mathrm{so(4)}$ maps $\mathrm{su(2)}\oplus\mathrm{su(2)}$ elements $X^+ + X^- = X^{+ i}\sigma_i + X^{- i}  \sigma_i$ into so(4) elements
\be
2 X = h(X^+,X^-)\,,
\ee
where
\be
X^\pm_i = \frac{1}{2} \epsilon_{0i}{}^{jk} X_{jk} \pm X_{0i}\,.
\ee
Inversion of the last equation gives
\bea
X_{ij} &=& \frac{1}{2} \epsilon_i{}^{jk}\left(X^+_k + X^-_k\right)\,, \\
X_{0i} &=& \frac{1}{2}\left(X^+_i - X^-_i\right)\,.
\eea
In the case of the group element $\gb'_{12}$, we have $X^\pm = -\phi^\pm J_3$, so
\bea
X_{ij} &=& -\frac{1}{4}(\phi^+ + \phi^-) \epsilon_{ij3}\,, \\
X_{0i} &=& -\frac{1}{4}(\phi^+ - \phi^-) \delta_{i3}\,,
\eea
and
\be
g'_{12} = \e^{2X} = \e^{-\frac{1}{2} (\phi^+ - \phi^-) J_{03}}\, \e^{-\frac{1}{2} (\phi^+ + \phi^-) J_{12}}\,.
\ee
Thus, if we define the angles $\alpha$ and $\alphat$ by
\be
g'_{12} = \e^{\alpha J_{03}}\, \e^{\alphat J_{12}}\,,
\ee
we find that
\be
\alpha = -(\phi^+ - \phi^-)/2\,,\qquad \alphat = -(\phi^+ + \phi^-)/2\,.
\ee
Therefore,
\be
j^+ \phi^+ + j^- \phi^- = - (j^+ - j^-)\alpha - (j^+ + j^-)\alphat\,.
\ee
Using that $j^\pm = |\gamma^\pm| j$ and $\gamma^- < 0$ for $\gamma < 1$ (see \cite{CF2}), we then arrive at
\be
j^+ \phi^+ + j^- \phi^- = -(\gamma^+ + \gamma^-) j\alpha - (\gamma^+ - \gamma^-) j\alphat\,.
\label{relationphialpha}
\ee

After the rotation \eq{gaugeforn1}--\eq{gaugeforn3}, the 4--vectors $\hat{N}_1$ and $\hat{N}_2$ become $\hat{N}'_1 = \hat{N}'_2 = (0,0,0,1)^T$ and the transformed bivectors equal
\be
X'_{12} = j \Uh\wedge \hat{N}'_1 = j J_{03} = X'_{21}\,.
\ee
Therefore,
\bea
\cos\alpha &=& \Uh\cdot g'_{12}\Uh\,, \\
\sin\alpha &=& \Uh\cdot g'_{12} \hat{N}'_2\,, \\
\cos\alphat &=& (0,1,0,0)^T\cdot g'_{12}(0,1,0,0)^T\,, \\
\sin\alphat &=& (0,1,0,0)^T\cdot g'_{12} (0,0,1,0)^T\,.
\eea
We know that $\Uh$ is not affected by the transformation \eq{gaugeforn1}--\eq{gaugeforn3}\footnote{This can be also checked directly by using the equation $g^- x^I\sigma^E_I (g^+)^{-1} = (R(\gb) x)^I\sigma^E_I$, which relates SU(2)$\times$SU(2) to SO(4) transformations.}, and $\hat{N}$ goes from $\hat{N}$ to $\hat{N}'$.
Hence, if we reverse this rotation, we obtain
\bea
\cos\alpha &=& \Uh\cdot g_{12}\Uh\,, \\
\sin\alpha &=& \Uh\cdot g_{12} \hat{N}_2\,, \\
\cos\alphat &=& R(n_1,n_1)(0,1,0,0)^T\cdot g_{12} R(n_2,n_2)(0,1,0,0)^T\,, \\
\sin\alphat &=& R(n_1,n_1)(0,1,0,0)^T\cdot g_{12} R(n_2,n_2)(0,0,1,0)^T\,,
\eea
where we recognize the vectors $V_{ij}$ and $W_{ij}$ on the right--hand side (see eqns.\ \eq{definitionVW}). Since $(0,1,0,0)^T$ and $(0,0,1,0)^T$ are orthogonal to the plane $X'_{12}$, the vectors $V_{12}$ and $W_{12}$ are orthogonal to $X_{12}$.

We conclude that $\alpha = \theta_{12}$ and $\alphat = \thetat_{12}$, 
and hence 
\be
j^+ \phi^+ + j^- \phi^- = -(\gamma^+ + \gamma^-) j\theta_{12} - (\gamma^+ - \gamma^-) j\thetat_{12}\,.
\label{relationphivartheta}
\ee
Thus, we obtain that
\be
A_{12} = \exp\left[-\irm (\gamma^+ + \gamma^-) j\theta_{12} - \irm(\gamma^+ - \gamma^-) j\thetat_{12}\right]\,,
\ee
at the saddle point, or more generally 
\be
A_{ij} = \exp\left[-\irm (\gamma^+ + \gamma^-) j\theta_{ij} - \irm(\gamma^+ - \gamma^-) j\thetat_{ij}\right]\,.
\ee
After taking into account determinant factors from the saddle point approximation, this yields eq.\ \eq{asymptoticsAv}.
The determinant factors can be easily obtained by expanding the logarithm of the amplitude around the stationary phase and using repeatedly (\ref{g}) and (\ref{h}).
These formula show that the action in the exponential is at quadratic order given by 
$ S = - \frac14 G^{a,b}_{+i,j} X^{+i}_{a}X^{+j}_{b} - \frac14 G^{a,b}_{- i,j} X^{-i}_{a}X^{-j}_{b} +O(X^{3})  $ with $g_{i}^{\pm} = \exp( X^{\pm i}_{a} \tau^{a}) \Omega_{i}^{\pm}$ 
and $\Omega^{\pm}_{i}$ denotes the spin connection which solves the stationary equations.
Eventually, we have a factor $(4\pi)^{12}$ coming from the stationary phase evaluation and a factor $(4\pi)^{-16}$ coming from the normalisation of the  measure on SU(2).
Because of the overall gauge invariance $g_{i}^{\pm}\to g^{\pm} g_{i}^{\pm}$  there are 8 integrations over SU(2) to be performed instead of 10.
Each SU(2) integration provides a factor $(4\pi)^{-2}$, because the
 normalised SU(2) measure is  
\be
\frac1{(4\pi)^{2}}
\left(\frac{2}{|X|} \sin\frac{|X|}{2}\right)^{2} \prod_{a} \d X^{a},\quad |X|< 2\pi\,.
\ee
\end{appendix}

\end{document}